\documentclass[structabstract]{aa}  
\usepackage{natbib}
\bibpunct{(}{)}{;}{a}{}{,} 
\usepackage{graphicx}
\usepackage{aalongtable}
\usepackage{txfonts}
\begin{document}
   \title{Variability in the  CoRoT photometry of three hot O-type stars\thanks{
            The CoRoT space mission was developed and is operated by the
            French space agency CNES, with participation of ESA's RSSD and
            Science Programmes, Austria, Belgium, Brazil, Germany and 
            Spain.}\fnmsep
            \thanks{Tables 2, 3 and 4 are only available in electronic form 
            at the CDS via anonymous ftp to 
            {\tt cdsarc.u-strasbg.fr (130.79.128.5)} or
            via {\tt http://cdsweb.u-strasbg.fr/cgi-bin/qcat?J/A+A/}.
}}

   \subtitle{HD~46223, HD~46150 and HD~46966}

   \author{R. Blomme\inst{\ref{inst1}}
           \and L. Mahy\inst{\ref{inst2}}
           \and C. Catala\inst{\ref{inst3}}
           \and J. Cuypers\inst{\ref{inst1}}
           \and E. Gosset\inst{\ref{inst2},}\thanks{Senior Research Associate F.R.S.-FNRS}
           \and M. Godart\inst{\ref{inst2}}
           \and J. Montalban\inst{\ref{inst2}}
           \and P. Ventura\inst{\ref{inst4}}
           \and G. Rauw\inst{\ref{inst2}}
           \and \\ T. Morel\inst{\ref{inst2}}
           \and P. Degroote\inst{\ref{inst5}}
           \and C. Aerts\inst{\ref{inst5},\ref{inst6}}
           \and A. Noels\inst{\ref{inst2}}
           \and E. Michel\inst{\ref{inst3}} 
           \and F. Baudin\inst{\ref{inst7}}
           \and A. Baglin\inst{\ref{inst3}}
           \and M. Auvergne\inst{\ref{inst3}}
           \and R. Samadi\inst{\ref{inst3}}
          }

   \institute{
              Royal Observatory of Belgium, Ringlaan 3, 1180 Brussel, Belgium;
              \email{Ronny.Blomme@oma.be}\label{inst1}
         \and
              Institut d'Astrophysique et de G\'eophysique, University of 
              Li\`ege, B\^at. B5C, All\'ee du 6 Ao\^ut 17, 4000 Li\`ege, 
              Belgium\label{inst2}
         \and
              LESIA, UMR 8109, Observatoire de Paris, 
              5 place Jules Janssen, 92195 Meudon Cedex,
              France\label{inst3}
         \and
              INAF, Osservatorio Astronomico di Roma, Via Frascati 33, 00040
              Monteporzio Catone (Roma), Italy\label{inst4}
         \and
              Instituut voor Sterrenkunde, K.U. Leuven, Celestijnenlaan 200D, 
              3001 Leuven, Belgium\label{inst5}
         \and
              Department of Astrophysics, IMAPP, University of Nijmegen, 
              PO Box 9010, 6500 GL Nijmegen, The Netherlands\label{inst6}
         \and
              Institut d'Astrophysique Spatiale (IAS), B\^atiment 121, 
              Universit\'e Paris-Sud, 91405, 
              Orsay Cedex, France\label{inst7}
             }

   \date{Received ; accepted }

 
  \abstract
{The detection of pulsational frequencies in stellar photometry is required
as input for asteroseismological modelling. The second short
run (SRa02) of the CoRoT mission has
provided photometric data of unprecedented quality and time-coverage
for a number of O-type stars.
}
{We analyse the CoRoT data corresponding to
three hot O-type stars, describing the properties
of their light curves and we search for pulsational frequencies, which we
then compare to theoretical model predictions.
}
{We determine the amplitude spectrum of the data, using the Lomb-Scargle
and a multifrequency HMM-like 
technique. Frequencies are extracted by prewhitening, and their
significance is evaluated under the assumption that the light curve is
dominated by red noise. We search for harmonics, linear combinations
and regular spacings among these frequencies.
We use simulations with the same time sampling as the data as a 
powerful tool to judge the significance of our results. 
From the theoretical point of view, we use
the MAD non-adiabatic pulsation code to determine the 
expected frequencies of 
excited modes.
}
{A substantial number of frequencies is listed, but none can be convincingly
identified as being connected to pulsations. The amplitude spectrum
is dominated by red noise.
Theoretical modelling shows that all three O-type stars can have excited 
modes but the relation between the theoretical frequencies and the
observed spectrum is not obvious. 
}
{The dominant red noise component in the hot O-type stars studied here 
clearly points to a different origin than the pulsations seen in 
cooler O stars. 
The physical cause of this red noise is unclear, but we
speculate on the possibility of sub-surface convection, granulation,
or stellar wind inhomogeneities being responsible.
}

   \keywords{stars: variables: general -- 
             stars: early-type -- 
             stars: individual: HD~46223 -- 
             stars: individual: HD~46150 -- 
             stars: individual: HD~46966 -- 
             stars: oscillations
               }
   \titlerunning{}

   \maketitle
%

\section{Introduction}

Even though the O-type stars are located in the 
Hertzsprung-Russell (HR) diagram inside a
zone where pulsations are expected, no asteroseismological
modelling of these stars could be performed prior to the
CoRoT 
\citep[Convection, Rotation and planetary Transits,][]{2006ESASP1306...33B, 2009A&A...506..411A}
mission.
Actually, few examples of variability have been identified. The
most obvious ones are the O9.5 V stars 
\object{$\zeta$ Oph} \citep{1997ApJ...481..406K}
and \object{HD 93521} \citep{1993A&A...279..148H, 2008A&A...487..659R} and 
their variability is
likely related to non-radial pulsations with periods of a few hours.
However, the majority of these detections were made spectroscopically.
The reason is that the amplitudes of pulsations are
too low to be measured from ground-based photometry.
Furthermore, these pulsations can be contaminated by variable
stellar winds.

The second short run (SRa02) of the CoRoT
satellite in the Asteroseismologic Channel
was partly devoted to the investigation of 
the photometric variability of O-type stars.
Pointing towards the anti-centre
of the Galaxy, this instrument observed objects belonging
to the young open cluster \object{NGC 2244}
inside the \object{Rosette nebula} and to the surrounding association
\object{Mon OB2}. The CoRoT data of three O-type stars
that are part of this run have been analysed in
previous papers: \object{HD 46149} by \citet{2010A&A...519A..38D},
\object{HD 47129} by \citet{2011A&A...525A.101M} and
\object{HD 46202} by \citet{2011A&A...527A.112B},
the latter paper containing the only forward modelling in terms of seismic
data of an O star so far.

In the present paper, we study the three remaining O-type stars that 
are part of the SRa02 run.
One of the targets is \object{HD~46223}, the hottest member of
NGC 2244. Situated at about 1.4--1.7 kpc \citep{2000A&A...358..553H},
this star was previously quoted in the literature as
having a spectral type
O5~V \citep{1956ApJS....2..389H, 1982A&A...107..252B}, 
O5~((f)) \citep{1974ApJ...193..113C},
O4~((f)) \citep{1995ApJ...454..151M}
and O4~V((f$^+$)) \citep{2009A&A...502..937M}.
The 9-year spectroscopic campaign by \citeauthor{2009A&A...502..937M} 
to investigate multiplicity 
did not reveal any significant variation of the radial
velocity related to the presence of a companion.
In addition, Chandra mosaic observations
of NGC 2244 \citep{2008ApJ...675..464W} detect no strong clustering
of X-ray sources around HD~46223, leading to the conclusion
that there is also no lower-mass neighbour present.
According to \citeauthor{2008ApJ...675..464W}, the small number
of close-by X-ray sources could indicate 
that this star is younger in comparison
to the population of the central part of the cluster.

\begin{table*}
\caption{CoRoT data and results from the analysis for the three stars.}
\label{table corot data}
\centering
\begin{tabular}{lllll}
\hline\hline
& Parameter & HD~46223 & HD~46150 & HD~46966 \\
\hline
\multicolumn{5}{l}{CoRoT data} \\
& number of raw data, $N_{\rm R}$ & 92687 & 92695 & 92688 \\
& number of non-flagged data, $N$ & 80403 & 80282 & 82273 \\
& duty cycle ($N/N_{\rm R}$, in \%) & 86.75 & 86.61 & 88.76 \\
& start date non-flagged data (HJD - 2\,450\,000)& 4748.488547 & 4748.485591 & 4748.488454 \\
& end date non-flagged data (HJD - 2\,450\,000) & 4782.819650 & 4782.819657 & 4782.805898 \\
& duration, $T$ (d)               & 34.331103   &   34.334066 &   34.317444 \\
& freq resolution (d$^{-1}$) & 0.029128    &   0.029126  &   0.029140  \\
\hline
\multicolumn{5}{l}{Spectral window} \\
& window peak (\%) at 2.0 d$^{-1}$    &  3.5  &  3.8  &  3.9 \\
& window peak (\%) at 4.0 d$^{-1}$    &  1.1  &  1.0  &  1.1 \\
\hline
\multicolumn{5}{l}{Stopping criterion: number of frequencies} \\
& AIC$_{\rm c}$ & 860 & 821 & 595 \\
& BIC           & 491 & 444 & 277 \\
& HQC           & 662 & 625 & 427 \\
\hline
\multicolumn{5}{l}{Number of red-noise significant frequencies} \\
&               & 59 & 50 & 53 \\
\hline
\multicolumn{5}{l}{Number of red-noise significant frequencies present in both halves of the observing run} \\
&               & 10 & 7 & 3 \\
\hline
\multicolumn{5}{l}{Red noise -- fit parameters} \\
& $\alpha_0$ (counts)    & $442 \pm 31$  & $430 \pm 65$  & $416 \pm 85$  \\
& $\tau$ (d)   & $0.09 \pm 0.01$ & $0.08 \pm 0.02$ & $0.17 \pm 0.05$ \\
& $\gamma$  & $0.96 \pm 0.02$  & $0.96 \pm 0.03$ & $0.91 \pm 0.02$ \\
\hline
\end{tabular}
\end{table*}

\begin{figure}
\resizebox{\hsize}{!}{\includegraphics[bb=20 260 555 685,clip]{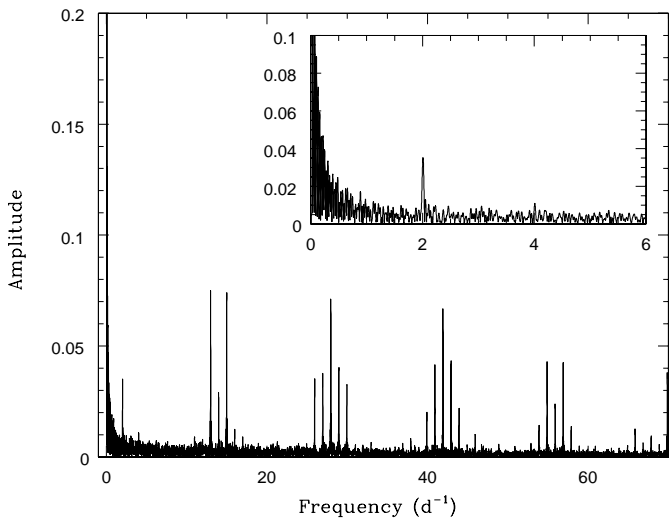}}
\caption{Spectral window of the HD~46223 observations.
The inset zooms in on the low-frequency region.
}
\label{fig spectral window}
\end{figure}

The next target is \object{HD 46150}, the second hottest star in 
NGC 2244.
Literature values for the spectral type are between
O5~V((f)) \citep{1990ApJ...364..626U, 2004ApJS..151..103M}
and O5.5~V((f)) \citep{1977ApJ...213..438C, 1980ApJ...242.1063G}.
The detailed study of \cite{2009A&A...502..937M}
confirms the O5.5~V((f)) spectral type. Radial velocity variations
have been detected in this star, but it is not clear if these are
due to binarity or motions in the stellar atmosphere
\cite[see][and references therein]{2009A&A...502..937M}.
\citeauthor{2009A&A...502..937M} monitored this star but could not
find any significant period in their data. 
They note the lack of spectral variability
over short timescales ($\sim$\,7 days).

The third target studied in the present paper is the late O-type
star \object{HD 46966}, belonging to the Mon OB2 association.
The spectral classification generally agrees on O8.5~V 
\citep{1977ApJ...214..759C, 1980ApJ...242.1063G, 2009A&A...502..937M} or
O8~V \citep{1999A&AS..137..521M, 2004ApJS..151..103M}.
Moreover, spectroscopic monitoring reveals no clues on the 
existence of a secondary component \citep{2009A&A...502..937M}.

These three stars are among the youngest
\citep[1--6 Myr,][]{2009MNRAS.394.2127B} and most massive observed
by the CoRoT satellite. They therefore have the potential to extend
our knowledge about pulsations into the higher-mass domain.
The CoRoT light curves are of unprecedented quality and
have a long observational time span (longer than
the rotational cycle). They thus
provide the ideal data set to reveal the 
presence or absence of such pulsations.

We first present, in Sect.~\ref{sect HD 46223}, the detailed 
frequency analysis of the HD~46223 CoRoT light curve.
In Sect.~\ref{sect HD 46150} and Sect.~\ref{sect HD 46966}, we
give the frequency analysis of HD~46150 and HD~46966 respectively.
Section~\ref{sect asteroseismological modelling} is devoted to 
determining the theoretically expected pulsation frequencies.
We discuss the 
results in Sect.~\ref{sect discussion} and in
Sect.~\ref{sect conclusions} we provide the conclusions of our research.

\section{HD~46223}
\label{sect HD 46223}

\subsection{CoRoT data}

The data we use here were obtained during the CoRoT second short run
(SRa02) made from 08 Oct to 12 Nov 2008. We start from the level
N2 data \citep{2007astro.ph..3354S}, which for each observation
list the time, 
flux, error on the flux and flagging information.
The flagging indicates whether these data are
contaminated by instrumental and environmental
conditions of the CoRoT satellite
such as, e.g., the South Atlantic Anomaly (SAA) and other Earth
orbit perturbations \citep[see][]{2009A&A...506..411A}.
In our analysis, we discard all flagged points, as well as those
few points that are listed with a negative flux error.

\subsection{Spectral window}

The spectral window for the HD~46223 observations
below 10~d$^{-1}$
(Fig.~\ref{fig spectral window}) 
is characterised
by a first peak at $f$ = 2.007~d$^{-1}$ and its double at about
$f$ = 4.011~d$^{-1}$. This is related to the passage of the satellite,
twice in a sidereal day, through the SAA. The relative amplitude
of these two peaks is 3.5~\% and 1.1~\%, respectively. In addition,
the gaps in the data set, due to the orbital period of the
satellite (6184 s), generate other structures with peaks around
$f$ = 13.972~d$^{-1}$ and their harmonics. 

The spectral window also
exhibits a peak at $f$ = 2699.76~d$^{-1}$, corresponding to
the sampling regularity of about 32.003~s. As a consequence,
the pseudo-Nyquist frequency is located at about 1350~d$^{-1}$. The
word ``pseudo'' emphasizes the fact that 
the sampling is not ideally regular because of the flagged data. The
duty cycle of the observations is 86.75 \%.
Because the data for all three stars were collected simultaneously
with the same instrument, the spectral window properties for the
other stars are very similar (see Table~\ref{table corot data}).

\subsection{Detrending}
\label{sect detrending}

In common with almost all CoRoT targets, there is a long-term trend 
visible in the data. This decreasing slope is likely due to the 
CCD ageing \citep{2009A&A...506..411A}. To remove this trend we 
divide the flux counts by the best-fit linear slope to the data. 

The final version of the CoRoT light curve covers $T\approx 34.3$~days.
The exact start and end times, duration,
number of points and frequency resolution
are listed in Table~\ref{table corot data}.
We do not convert the fluxes to magnitudes.

\begin{figure*}
\centering
\includegraphics[width=17cm]{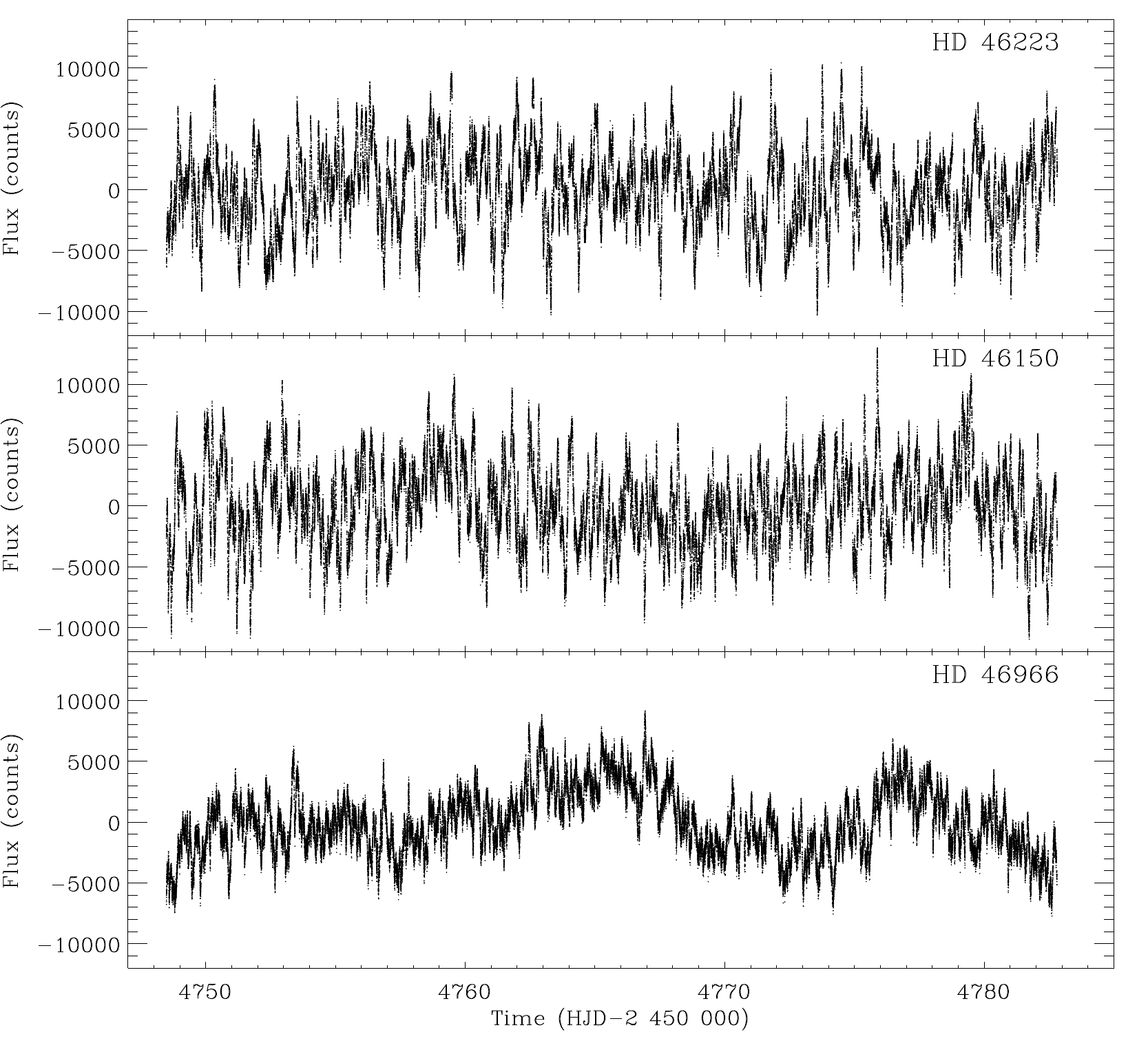}
\caption{Detrended CoRoT light curves, covering the full 34.3 days
of the observing run. Each observation is represented by a dot.
Flagged data are not plotted.
{\em Top:} HD~46223, {\em middle:} HD~46150, {\em bottom:} HD~46966.
}
\label{fig light curves}
\end{figure*}

\begin{figure*}
\centering
\includegraphics[width=17cm]{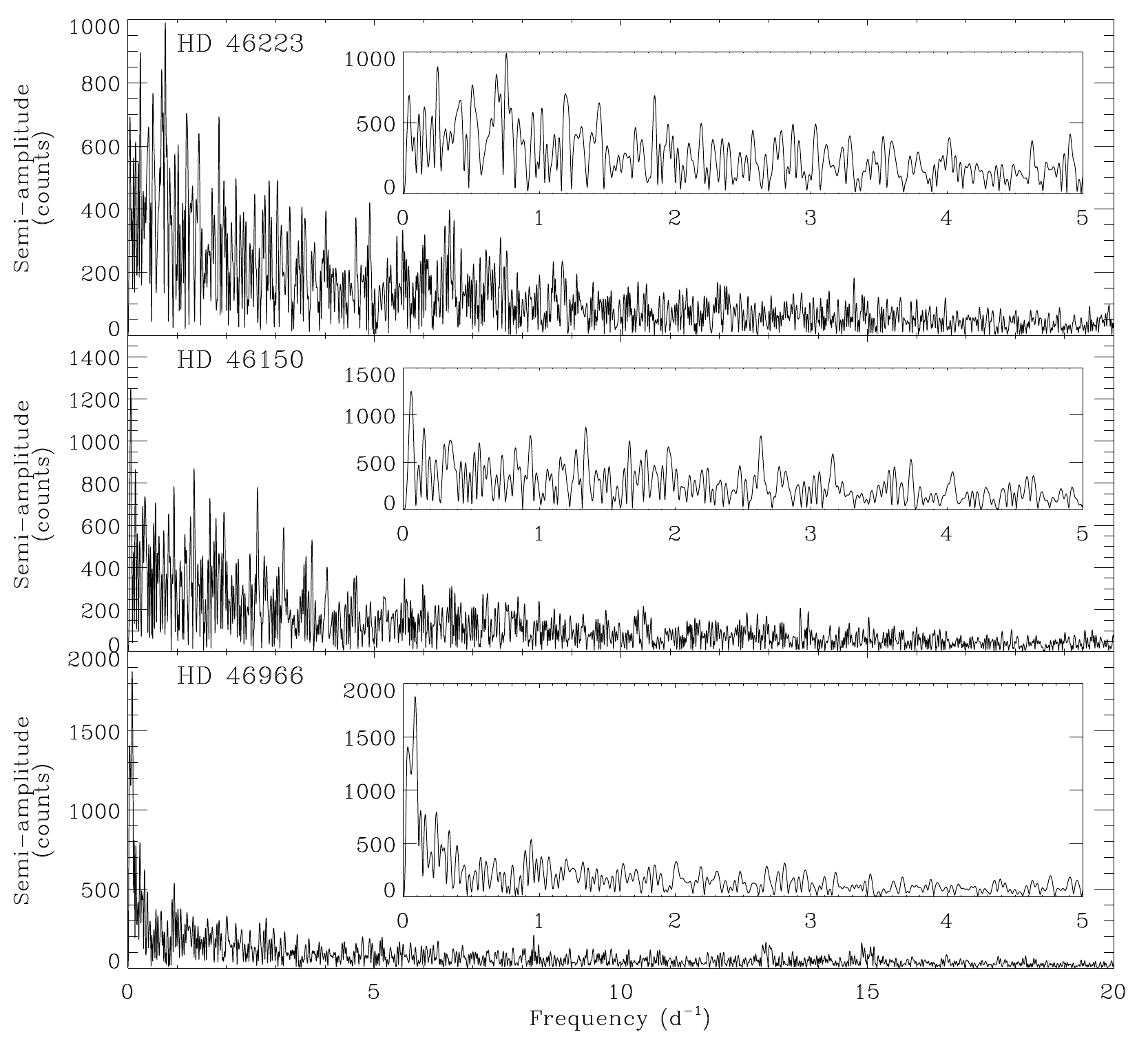}
\caption{Amplitude spectrum of the CoRoT light curves
computed by the Lomb-Scargle method.
The insets show a zoom-in of the amplitude spectrum on the 
low-frequencies domain.
{\em Top:} HD~46223, {\em middle:} HD~46150, {\em bottom:} HD~46966.
}
\label{fig spectrum}
\end{figure*}

\subsection{Frequency-by-frequency prewhitening}
\label{sect prewhitening}

The range of variations observed in the HD~46223 CoRoT light curve 
(Fig.~\ref{fig light curves}, top)
is of order 16,000 counts, corresponding to 8~mmag.
No clear pattern in the variations is visible.
In order to better understand the variability present in the
CoRoT light curve,
we apply a Fourier analysis based 
on the Lomb-Scargle method \citep{1976Ap&SS..39..447L, 1982ApJ...263..835S}.

The amplitude spectrum (Fig.~\ref{fig spectrum}, top) exhibits many 
peaks
and appears to be noisy. The highest peak is located
at about $f$ = 0.75~d$^{-1}$ and the second one at about $f$ = 0.25~d$^{-1}$,
but no high-significance peaks are clearly detected, nor does there
appear to be systematics, such as a constant spacing between peaks.
A plot of the spectrum up to the pseudo-Nyquist frequency shows
very little power at high frequencies. In the further analysis,
we therefore limit the frequency domain to 0--100~d$^{-1}$.

For a more detailed analysis, we perform a
traditional prewhitening. We start by subtracting the average from the
fluxes. We then determine the amplitude spectrum, using the 
Lomb-Scargle method. We select the frequency
corresponding to the highest peak
in the amplitude spectrum and fit a sine function 
($A_j \sin (2\pi f_j t_i + \phi_j)$)
to the fluxes, as a function of time ($t_i$). 
In the fitting procedure, the semi-amplitude ($A_j$) and 
phase ($\phi_j$) are
free parameters, while the frequency ($f_j$) is allowed to vary within 
a range of $\pm 0.1/T$ from its peak 
value\footnote{The $\pm 0.1/T$ value turns out to be quite close to 
      (slightly higher than) the formal
      errors we get on the frequencies (see 
      Eq.~\ref{eq errors f}--\ref{eq errors phi}
      and Tables~\ref{table freq HD 46223}, 3 and 4).}
(where $1/T$ is the frequency resolution
and $0.1/T$ the step size in our frequency grid).
The fitted sine function is then subtracted from the data and 
the amplitude spectrum is re-computed from these prewhitened data.
This procedure is continued until we reach the noise level in the 
observations.
After the prewhitening, the observed variability, whatever its nature,
is thus described (down to the noise level) as a
sum of terms, with each term being a sine function at a particular frequency.

\subsection{Stopping criteria}
\label{sect stopping criteria}

To judge when we have reached the noise level, we explore
three stopping criteria that have been discussed in the 
literature \citep{HQC,HANNAN,2007MNRAS.377L..74L}. The
information content of the fit is given
by:
\begin{eqnarray}
c_N(m) & = & -2 \ln (L_N(m)) \nonumber \\
& + & \left\{ 
      \begin{array}{l@{\quad}l}\vspace{0.7em}
    2m + \frac{\displaystyle 2m(m+1)}{\displaystyle (N-m-1)} &  {\rm Akaike, corrected~(AIC_c)}\\\vspace{0.7em}
    m\; \ln(N) & {\rm Bayesian~(BIC)} \\
    2m\; \ln(\ln(N)) & {\rm Hannan-Quinn~(HQC)} \\
       \end{array}
\right. ,
\label{eq stop criteria}
\end{eqnarray}
where $N$ is the number of observations, $m$ is the number of 
parameters and $L_N(m)$ is the likelihood function. 
To within a constant,
$-2 \ln (L_N(m))$ can be written as 
\citep[][Sect. 15.1]{1992nrfa.book.....P}:
\begin{equation}
\sum_{i=1}^N \frac{(y_i - F(t_i,\mathbf{\Theta}_m))^2}{\sigma_i^2} ,
\end{equation}
where $y_i$ are the fluxes observed at times $t_i$, and $\mathbf{\Theta}_m$
is the vector containing the $m$ fit-parameters.
For the value of $\sigma_i$ we take the error bars listed with 
each observed flux.
We stop the prewhitening process when 
$c_N(m)$
has reached a minimum. This is very similar to least-squares
minimization, but the additional terms in 
Eq.~\ref{eq stop criteria}
correct for the fact
that a model fit improves with an increasing number of parameters.
We set $m = 3k$, counting 
3 parameters (frequency, amplitude, phase) for 
each of the $k$ sine functions we fit.
Note that we subtract the average from the observed fluxes before 
starting the prewhitening procedure (Sect.~\ref{sect prewhitening}),
which could lead to a bias.

Applying this to the HD~46223 data, we find that
the minimum of AIC$_{\rm c}$ is reached at 860 terms, BIC at 491 and 
HQC at 662.
To judge the reliability of these stopping criteria, we
run a number of 
simulations
where we construct artificial light curves
with a given number of sine functions. 
We use the same time sampling as the observations and
attribute to it the same noise level and flagging
(this approach is used in all simulated light curves discussed in this paper). 
For the amplitudes and frequencies
we take either values from the analysis of the observations, 
or randomly drawn
values distributed according to the statistics of the observed
spectrum. 
Noise is added
to the simulation, using a Gaussian random variable (with mean zero
and standard deviation unity) multiplied
by the listed error on the observed flux.
We then apply our prewhitening procedure to the simulations and
compare the results of the stopping criteria with the known number of
terms we introduced. It turns out that BIC gives the best value,
which is always within a factor 1.5 of the correct result.
\citet{2009A&A...506..111D} also used BIC in preference over
the (uncorrected) AIC as a stopping criterion in their study
of the $\beta$~Cep star \object{HD~180642}.

As a further test, we also redo the analysis on the observed data
after rebinning over a number of
points (up to 32 points, i.e. 17\,min), 
but find approximately the same results. This is consistent
with the fact that there is little power at higher frequencies.
Based on the above,
we take the BIC rounded-off value of 500 terms
as the stopping criterion.
The final 500 frequencies and corresponding semi-amplitudes and phases
are listed in Table~\ref{table freq HD 46223}
(the full Table~\ref{table freq HD 46223} is only available in
electronic form at CDS).
The phases are defined with respect to $t=0$ corresponding to the first
non-flagged data point (as listed in Table~\ref{table corot data}).

New frequencies found during the prewhitening procedure can be quite
close to previously found frequencies. The resolution criterion
for this is that frequencies should be separated by at least $1.5/T$
to be considered unique \citep{1978Ap&SS..56..285L}. Applying
this to our list of 500 frequencies,
we find that 276 of them are unique.

\begin{table*}
\caption{List of sine function terms for HD 46223.}
\label{table freq HD 46223}
\begin{tabular}{rrrrrrrrrcc}
\hline
\hline
\multicolumn{1}{c}{ID} & \multicolumn{1}{c}{Freq.} & \multicolumn{1}{c}{Freq.} & \multicolumn{1}{c}{Semi-ampl.} & \multicolumn{1}{c}{Semi-ampl.} & \multicolumn{1}{c}{Phase} & \multicolumn{1}{c}{Phase} & \multicolumn{1}{c}{Standard} & \multicolumn{1}{c}{Sig.} & \multicolumn{1}{c}{Red-} & \multicolumn{1}{c}{Half-} \\
 &  & \multicolumn{1}{c}{Error} & & \multicolumn{1}{c}{Error} & & \multicolumn{1}{c}{Error} & \multicolumn{1}{c}{Deviation} & \multicolumn{1}{c}{Red-} & \multicolumn{1}{c}{noise} & \multicolumn{1}{c}{run} \\
 & \multicolumn{1}{c}{(d$^{-1}$)} & \multicolumn{1}{c}{(d$^{-1}$)} & \multicolumn{1}{c}{(counts)} & \multicolumn{1}{c}{(counts)} & \multicolumn{1}{c}{(rad)} & \multicolumn{1}{c}{(rad)} & \multicolumn{1}{c}{(counts)} & \multicolumn{1}{c}{noise} & \multicolumn{1}{c}{Significant} & \multicolumn{1}{c}{Test} \\
\hline
   1  & 0.7548  & 0.0050  & 989.095  & 305.732 & -2.500  & 0.309  & 3296.681  & 1.0000 &  &  * \\
   2  & 0.2493  & 0.0068  & 882.119  & 371.151 & -1.830  & 0.421  & 3221.288  & 1.0000 &  &  * \\
   3  & 0.5102  & 0.0067  & 772.225  & 323.271 &  1.476  & 0.419  & 3160.377  & 1.0000 &  &  * \\
   4  & 0.6858  & 0.0063  & 752.212  & 296.426 & -0.497  & 0.394  & 3112.720  & 1.0000 &  &  * \\
   5  & 1.1944  & 0.0056  & 699.589  & 244.181 &  0.676  & 0.349  & 3067.088  & 1.0000 &  &  * \\
   6  & 1.8456  & 0.0046  & 691.462  & 199.852 & -2.537  & 0.289  & 3026.889  & 1.0000 &  &  * \\
   7  & 0.0395  & 0.0091  & 681.390  & 385.046 & -1.697  & 0.565  & 2986.975  & 1.0000 &  &  * \\
   8  & 0.4251  & 0.0077  & 650.217  & 312.906 & -0.054  & 0.481  & 2948.151  & 1.0000 &  &  * \\
   9  & 0.9477  & 0.0064  & 636.137  & 251.785 & -1.726  & 0.396  & 2912.468  & 1.0000 &  &  * \\
  10  & 1.4370  & 0.0053  & 642.207  & 212.699 & -1.822  & 0.331  & 2877.663  & 1.0000 &  &  * \\
\multicolumn{1}{c}{$\cdots$} & \multicolumn{1}{c}{$\cdots$} & \multicolumn{1}{c}{$\cdots$} & 
   \multicolumn{1}{c}{$\cdots$} & \multicolumn{1}{c}{$\cdots$} & \multicolumn{1}{c}{$\cdots$} & 
   \multicolumn{1}{c}{$\cdots$} & \multicolumn{1}{c}{$\cdots$} & \multicolumn{1}{c}{$\cdots$} & 
   \multicolumn{1}{c}{$\cdots$} & \multicolumn{1}{c}{$\cdots$} \\
  59 &  7.5519  & 0.0027  & 303.826  &  51.957 &  0.500  & 0.171  & 1913.541  & 0.0080 & R &   \\
\multicolumn{1}{c}{$\cdots$} & \multicolumn{1}{c}{$\cdots$} & \multicolumn{1}{c}{$\cdots$} & 
   \multicolumn{1}{c}{$\cdots$} & \multicolumn{1}{c}{$\cdots$} & \multicolumn{1}{c}{$\cdots$} & 
   \multicolumn{1}{c}{$\cdots$} & \multicolumn{1}{c}{$\cdots$} & \multicolumn{1}{c}{$\cdots$} & 
   \multicolumn{1}{c}{$\cdots$} & \multicolumn{1}{c}{$\cdots$} \\
\hline
\end{tabular}
\tablefoot{This is an excerpt of Table~2. The full table is only available in
electronic form at the CDS via anonymous ftp to 
{\tt cdsarc.u-strasbg.fr (130.79.128.5)} or
via {\tt http://cdsweb.u-strasbg.fr/cgi-bin/qcat?J/A+A/}.
We list the first ten terms as well as the first term that is 
significant under the assumption of red noise (indicated with 'R').
Terms that are found in both halves of the observing run are indicated
with '*' in the last column.
}
\end{table*}

The canonical errors estimated on these frequencies ($f$), 
and their corresponding
semi-amplitudes ($A_f$) and phases ($\phi_f$) are given by the
expressions of \cite{1971AJ.....76..544L} and 
\cite{1999DSSN...13...28M}:
\begin{eqnarray}
\epsilon(f) & = & \sqrt{\frac{6}{N}} \: \frac{1}{\pi T} \: \frac{\sigma_{f}}{A_f}  \label{eq errors f} \\
\epsilon(A_f) & = & \sqrt{\frac{2}{N}} \: \sigma_{f} \label{eq errors A} \\
\epsilon(\phi_f) & = & \sqrt{\frac{2}{N}} \: \frac{\sigma_{f}}{A_f} ,
 \label{eq errors phi}
\end{eqnarray}
with $\sigma_f$ the standard 
deviation
on the light curve at the current stage of prewhitening.
These formal errors are also listed in
Table~\ref{table freq HD 46223}.

\subsection{Multifrequency fitting}
\label{sect multifrequency fitting}

The Lomb-Scargle based iterative prewhitening consists in
a one-by-one determination
of the sine functions and in their removal from the light curve. In a gapped
or unevenly sampled data set,
the height of peaks is dependent on the height of other peaks in the
periodogram.
A more general procedure should therefore determine all sine function terms
simultaneously.

\defcitealias{1985A&AS...59...63H}{HMM}
In order to be more reliable in the detection of the frequencies, we also 
use a multifrequency algorithm introduced by 
\citet[][their Eq. A13 to A19]{2001MNRAS.327..435G} 
and developed from the 
\citet[][hereafter \citetalias{1985A&AS...59...63H}]{1985A&AS...59...63H}
method.
We note that this multifrequency algorithm is not based on the 
Lomb-Scargle technique. However, in the case of CoRoT data, we find
that the amplitude spectrum computed from Lomb-Scargle and from
HMM are in very good agreement.
The multifrequency algorithm takes into account the 
mutual influence of peaks by fitting them all together, and is therefore
in principle an improvement on the one-by-one prewhitening procedure used
in Sect.~\ref{sect prewhitening}.
Similarly to the prewhitening procedure, the present method refines the
value of each frequency (allowing it to vary within the natural width
of the individual peaks).

We apply this method to the non-flagged data of the HD\,46223 CoRoT light 
curves but the great number of data points involves an excessive computation 
time. Accordingly, we have to deal with a limited number of terms at 
the same time. We find that the values of the 
frequencies fitted by the multifrequency programme are very similar to those 
determined by the prewhitening procedure and only small differences in 
the amplitudes are found. We therefore do not list these results.

\subsection{Significance levels}

Although the quality of the CoRoT curve is unparalleled, it is
still important to check whether peaks detected in the amplitude
spectrum might result from a random variation rather
than representing a periodic signal. To do this, we use
the statistical
criterion presented by \citet{Gosset2007} and quoted in 
\citet{2011A&A...525A.101M}.
The probability that at least one of the semi-amplitudes of an extended set of 
frequencies exceeds a threshold $z$ under the
null hypothesis of a stochastic process of variance $\sigma_f^2$
is given by
\begin{equation}
{\rm Prob}[Z_{\rm max} > z] = 1-e^{\displaystyle -e^{\displaystyle -0.93 z + \ln(0.8 N)}}
\label{eq significance}
\end{equation}
where $Z_{\rm max} = {\rm max}_{f<f_{\rm Ny}} Z(f)$,
$N$ is the number of
points in the data set and $Z(f) = A_f^2 N /(4\sigma_f^2)$ with $A_f$ the 
semi-amplitude of the frequency $f$ and $\sigma_f$ the standard deviation
on the light curve at the current stage of prewhitening.

We also explore a different version of this criterion,
which  uses \citep{2009A&A...506..471D}:
\begin{equation}
{\rm Prob}[Z_{\rm max} > z] = 1-(1-\exp(-z))^{N_i} ,
\end{equation}
where $N_i$ is the number of independent frequencies.
\citeauthor{2009A&A...506..471D} show that the CoRoT data are sufficiently
well equally-spaced that we can take $N_i = 2N$.
In our work, the first criterion (given by Eq.~\ref{eq significance})
turns out to be slightly stricter,
and we therefore discuss only results
obtained with that one.

In applying this criterion, 
the simplest approach is to assume white noise, i.e. 
$\sigma_f$ is independent of frequency and its value is 
given by the standard deviation 
on the light curve at the current stage of prewhitening.
Under that assumption all 500 sine function terms are highly significant.

\begin{figure}
\resizebox{\hsize}{!}{\includegraphics{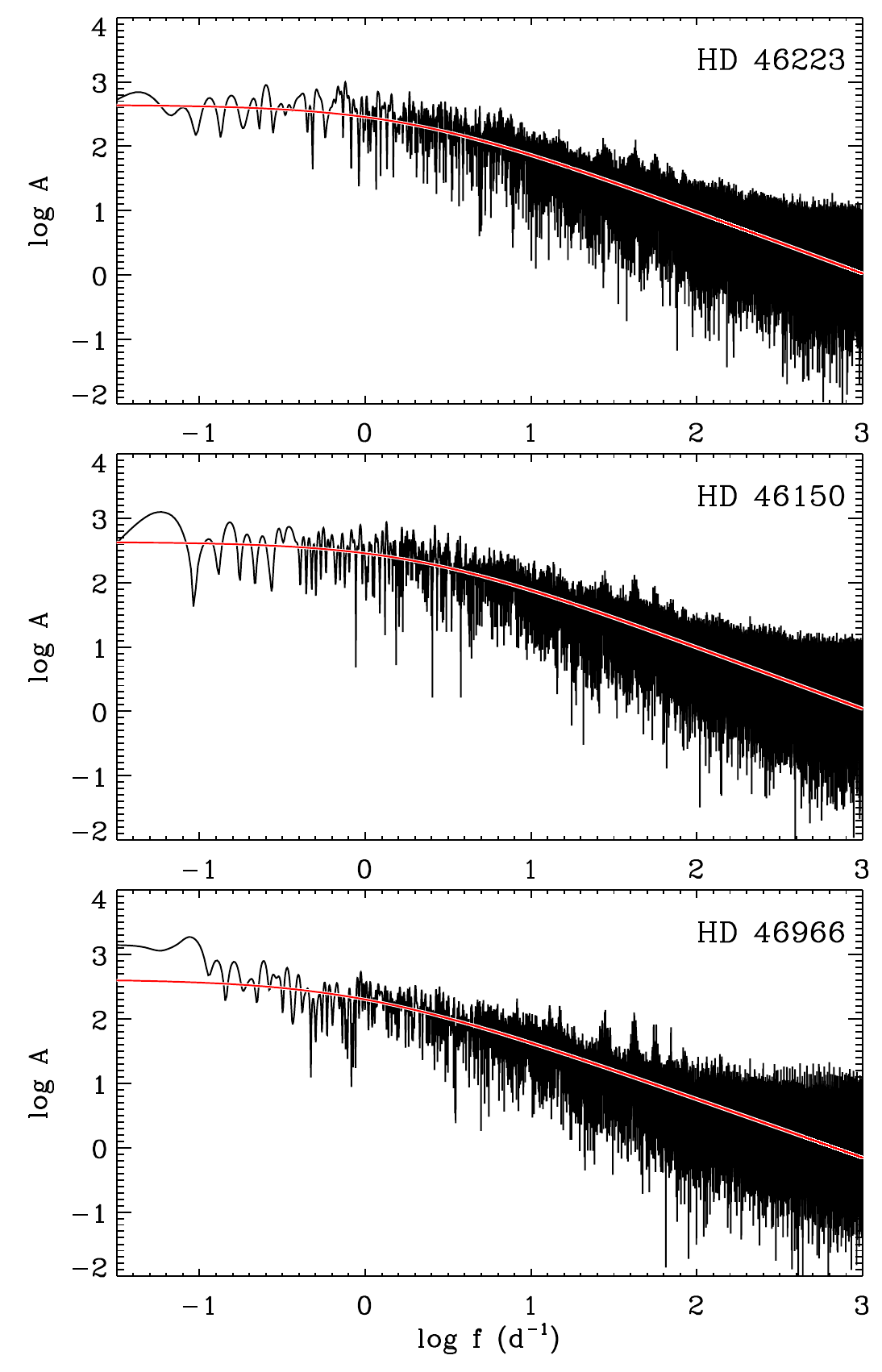}}
\caption{Fit-function (Eq.~\ref{eq fit function}, red line) to the 
amplitude spectra as a function of frequency, in a log--log plot.
{\em Top:} HD~46223, {\em middle:} HD~46150, {\em bottom:} HD~46966.
}
\label{fig red noise}
\end{figure}

\begin{figure}
\resizebox{\hsize}{!}{\includegraphics{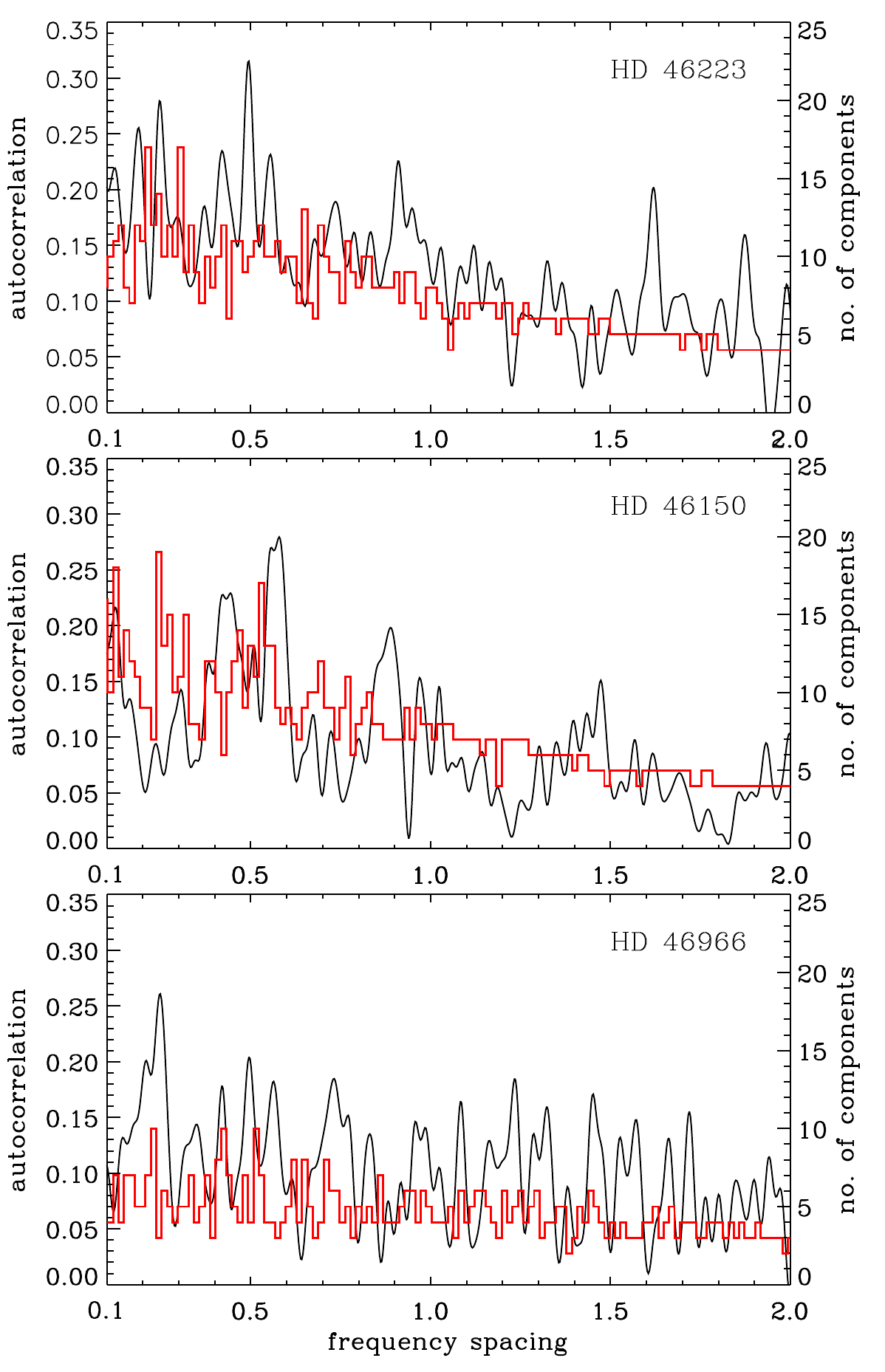}}
\caption{Autocorrelation functions (black curve)
of the power periodograms and number
of components in a chain with a given frequency spacing (red curve). 
{\em Top:} HD~46223, {\em middle:} HD~46150, {\em bottom:} HD~46966.
}
\label{fig spacing}
\end{figure}

\subsection{Red noise}

It is quite clear from the amplitude spectrum
(Fig.~\ref{fig spectrum}) that the power increases substantially
towards lower frequencies. This strongly suggests the presence of red 
noise. We stress that the word ``noise" does not imply an instrumental
origin but rather points to predominantly stochastic behaviour: 
in Sect.~\ref{sect pulsations and red noise}
we discuss the possible physical cause of the red noise.

In the present
section, we determine the observed properties of this red noise.
\citet{2002A&A...394..625S} suggest to describe
the amplitude spectrum $\alpha(f)$ by using the following fit-function:
\begin{equation}
\alpha(f)=\frac{\alpha_0}{1+(2\pi \tau f)^\gamma}
\label{eq fit function}
\end{equation}
with $\alpha_0$ the scaling factor, $\gamma$
the slope of the linear part (in a log--log plot) and $\tau$ an estimation
of the mean duration of the dominant structures in the light
curve. We fit this profile to the amplitude spectrum of
HD~46223 (Fig.~\ref{fig red noise}, top)
in order to quantify the
information about the red noise contained in the periodogram.
We apply a least-squares fit to the logarithm of both 
frequency and semi-amplitude,
limiting the frequencies used to 0--100~d$^{-1}$.
The resulting fit parameters 
and their 1-sigma error bars
are listed in Table~\ref{table corot data}.

To re-evaluate the significance of the peaks under the hypothesis
of red noise, we need to determine its $\sigma_{f,{\rm red}}$. We
do so by first converting the fit-function for amplitude
(Eq.~\ref{eq fit function})
into power. We then re-scale that function,
so that its integral over frequency
(up to the pseudo-Nyquist frequency of $\sim$ 1350 d$^{-1}$) is the same
as the variance on the flux multiplied by the pseudo-Nyquist frequency.

We then redo the prewhitening procedure and for each frequency
we determine
the significance of the peak by using Eq.~\ref{eq significance}
with $Z(f) = A_f^2 N /(4\sigma_{f,{\rm red}}^2)$.
It is important to realise that in doing so, we still select the highest
peak at each step in the prewhitening process, 
not the one that is most significant under red-noise conditions.
If we selected the most significant peak, 
it would correspond to a small-amplitude contribution to the
light curve. In the prewhitening procedure (Sect.~\ref{sect prewhitening})
we then need to fit a sine function with this frequency to the light
curve, and subtract it. The fitting procedure
will not succeed, because in attempting to fit only a minor 
contribution to the light curve, it will be led astray
by the much higher-amplitude
contributions (which have not been removed yet).
It will therefore give an incorrect 
result, or will fail to converge.

Applying this procedure,
we find 59 significant sine functions 
with a significance level less than
0.01; we indicate them with an 'R' in
Table~\ref{table freq HD 46223}.

\subsection{Split in two halves}
\label{sect split in two halves}

We next
apply an additional method to determine
the reliability of the sine function terms we found.
We split the data set into two halves, and redo the analysis on each
half separately. We then check how many of the original terms
can also be found in both halves. We consider two frequencies 
($f_1$ and $f_2$) to be
the same if 
$|f_1 - f_2| < \sqrt{\epsilon^2(f_1) + \epsilon^2(f_2)}$, where the
errors are defined in Eq.~\ref{eq errors f}. 
Simulations with artificial data confirm
that this is the appropriate criterion to use.
We could of course relax it by considering equality to within
2 or 3 times the 1-sigma error bar, but this would considerably
increase the number of false coincidences.

If we first look at the 
terms determined assuming white noise,
we find that,
of the 500 terms extracted 
from the full run, 202 are present in both halves.
This is substantially less than the 282 terms we find if we construct
an artificial, noisy, light curve from the first 100 observed 
terms of HD~46223. 
Other simulations with frequencies drawn
randomly according to the observed distribution also find a 
significantly larger number of terms in both halves.

If we limit ourselves to the 59 red-noise significant terms
listed in Table~\ref{table freq HD 46223}, we find that only 10
are present in both halves of the data set.

\subsection{Frequency combinations and spacings}
\label{sect combinations}

We next search for the presence of low-order harmonics (up to order 10)
in the list of 500 frequencies.
With the usual criterion that frequencies
should be equal to within the
square root of the sum of the squares of the
individual 1-sigma errors, we find 983 such harmonics.
However, simulations as in Sect.~\ref{sect stopping criteria}
show a factor 2--3 more harmonics. Our findings must therefore
be attributed to coincidences, not to true harmonics.

A similar search for linear combinations
of the type $f_1 + f_2 = f_3$ among the
500 frequencies reveals 45,412 results. But, again, simulations
show these to be spurious.
If we limit our search to frequencies significant under the red-noise
hypothesis, we find no harmonics and only one linear combination.
The reason we find so few of these is
that most of our red-noise frequencies are between 10 and 20 d$^{-1}$;
harmonics and sums of these are then outside the 10 -- 20 d$^{-1}$ range.
Because of the limited
value of these results, we do not list them.

\begin{figure*}
\centering
\includegraphics[width=17cm]{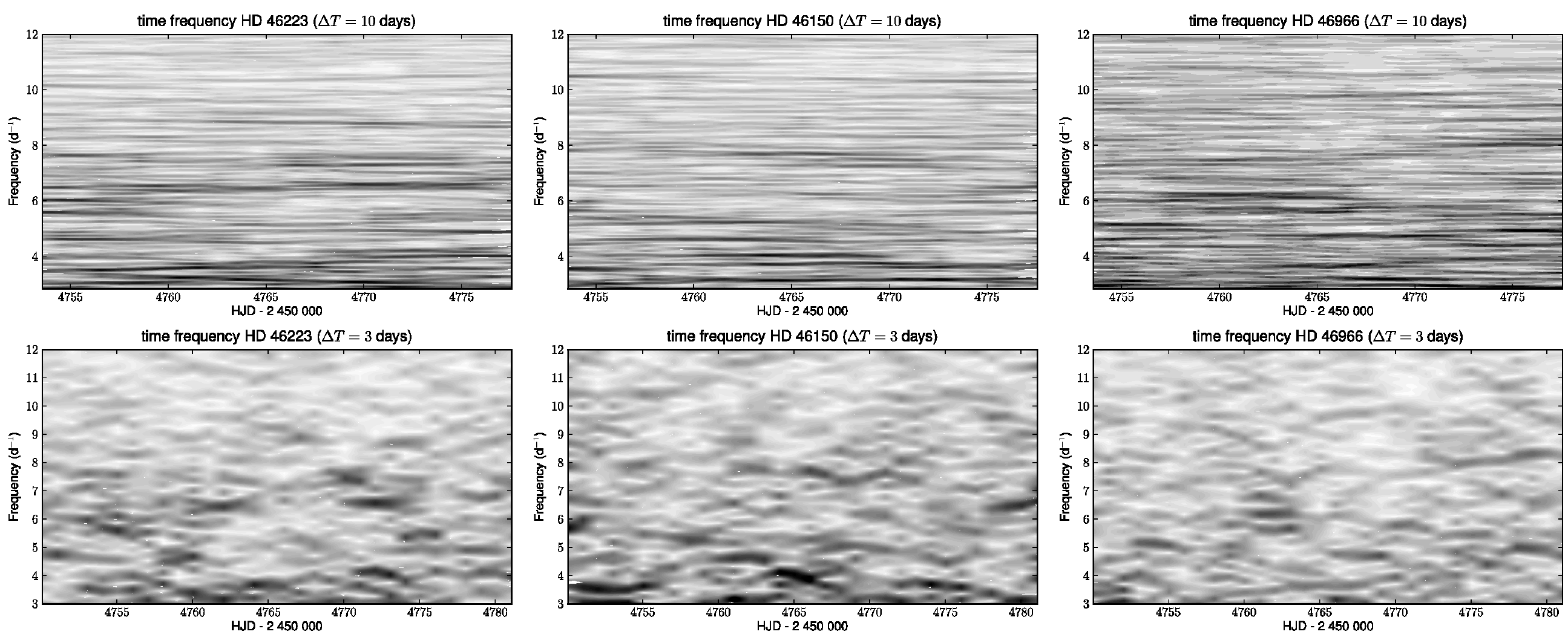}
\caption{Time-frequency analysis of the light curves. The top row uses a 
sliding window of 10 days, the bottom row one of 3 days.
{\em Left panels:} HD~46223, {\em middle:} HD~46150, 
{\em right:} HD~46966.
The grey scale indicates the semi-amplitude of the
frequencies. It is clear that the frequencies do not have long lifetimes.}
\label{fig time-frequency analysis}
\end{figure*}

In addition to these linear combinations, we investigate the
frequency spacing for HD~46223. 
\citet{2010A&A...519A..38D} found such spacing for the O8 V star HD~46149.
Following their work, we look at both the autocorrelation function
of the power periodogram and the number of components we find in
a chain ($f_0$, $f_0+\Delta f$, $f_0+2\Delta f$, ...) with an assumed
frequency spacing $\Delta f$. 
In both the chain and autocorrelation test, we start from the 500 frequencies,
but eliminate those outside the 3.0--12.0~d$^{-1}$ domain.
The lower limit is taken to avoid
the stronger peaks, the upper limit to avoid effects from the
satellite orbital period (13.972 d$^{-1}$). To determine the number 
of components in a chain at a given frequency, we take that
$f_0+n\Delta f$ equals a listed frequency if both
differ by less than the step size
we use in the figure, which is 0.015 d$^{-1}$.
For $\Delta f$, we consider only the range of 0.1 to 2.0 d$^{-1}$, as this
is where both the autocorrelation and chain-test function have their
largest values.

Fig.~\ref{fig spacing} (top) shows the results for HD~46223. The 
autocorrelation function has a peak at $\Delta f$ = 0.5~d$^{-1}$,
but the number of components does not confirm this peak. We
therefore conclude that we do not detect significant frequency
spacings in HD~46223.

\subsection{Time-frequency analysis}
\label{section time-frequency analysis}

We also perform 
a time-frequency analysis (Fig.~\ref{fig time-frequency analysis}) 
in order to detect variations in the amplitude of frequencies or in the 
frequencies themselves as a function of time.

For this purpose, a sliding window with lengths of 10 and 3
days is applied to the light curve and shifted with a step of
one day. Then we apply a Fourier analysis on each frame by
using the HMM method 
\citep[for further details on this 
time-frequency analysis method, see][]{2009A&A...506...95H}.
A plot of all frequencies in the 3--12 d$^{-1}$ range detected on a sliding 
window of 10 and 3 days is shown in Fig.~\ref{fig time-frequency analysis}.
This figure confirms what we already found from the test where we split
the data into two halves. The frequencies are clearly not stable over the
$\sim$~34-day timescale of the observation. The figures do not
show many frequencies that have a lifetime longer than the duration
of the sliding window. This behaviour is compatible
with the dominant presence of red noise.

The limited lifetime also calls into question the applicability of the
$1.5/T$ criterion \citep{1978Ap&SS..56..285L} to resolve frequencies
(see Sect.~\ref{sect stopping criteria}). A limited lifetime translates
into a broadening of the frequency. This will reduce the number of
unique frequencies found in our analysis.

\section{HD~46150}
\label{sect HD 46150}

For the analysis of HD~46150, we apply a similar detrending as for HD~46223.
The range of variations observed in the HD~46150 
CoRoT light curve 
(Fig.~\ref{fig light curves}, middle)
is of order 16,000 counts, corresponding to 8 mmag.
Further details are listed in Table~\ref{table corot data}.
No clear pattern in the variations is visible.
The amplitude spectrum (Fig.~\ref{fig spectrum}, middle) is noisy
and has its highest peak at 0.055~d$^{-1}$. Because there is little
power at high frequencies, we limit the frequency domain to 0--100~d$^{-1}$
in the further analysis.

\begin{figure*}
\sidecaption
\includegraphics[width=12cm]{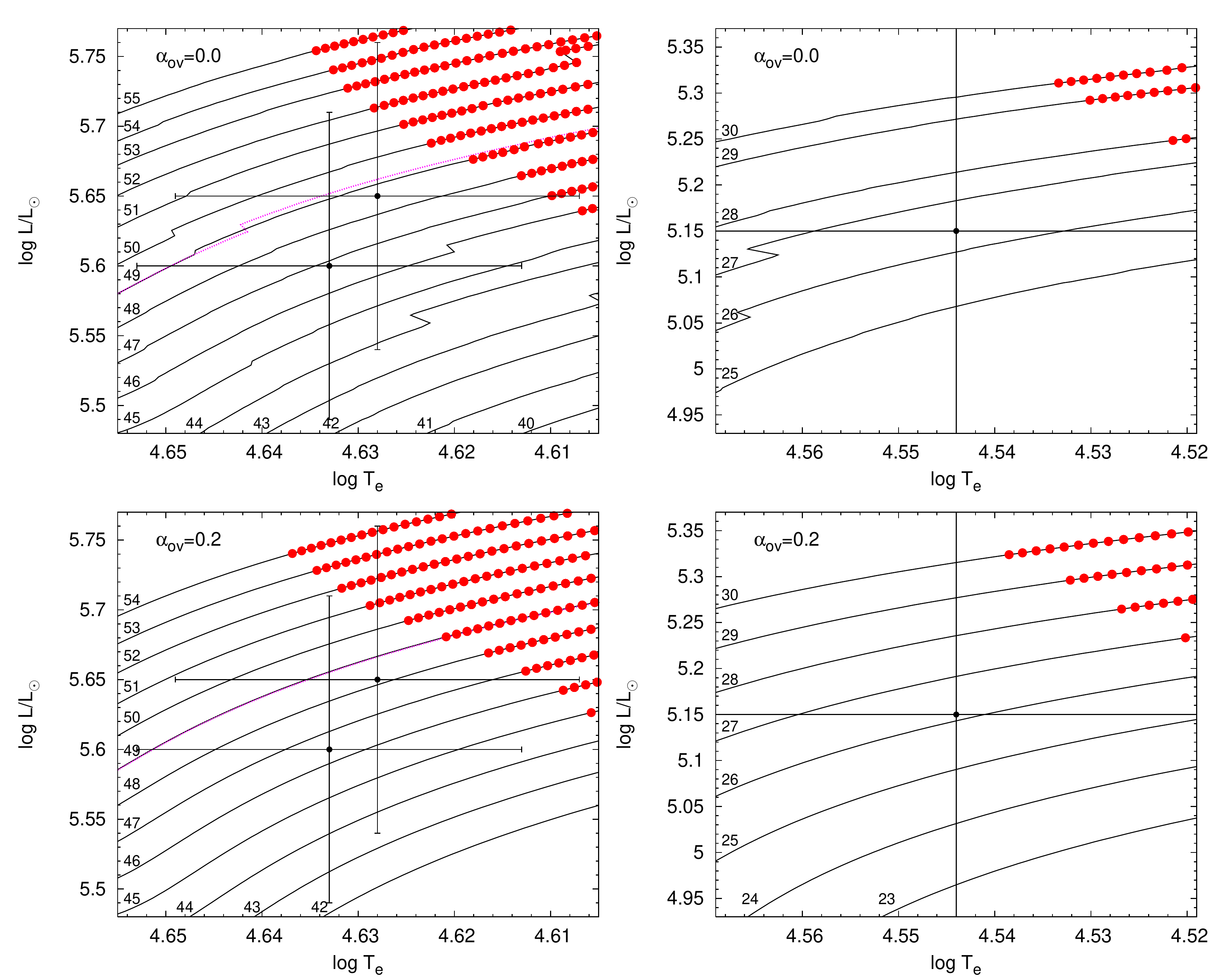}
\caption{Evolutionary tracks relevant for our three stars. 
The models in the bottom figures include overshooting, the ones at
the top do not.
{\it Left:} Tracks of 40 to 55\,$M_{\sun}$ computed with a mass-loss rate of 
$6.0 \times 10^{-8}$ $M_{\sun} {\rm yr}^{-1}$. The error boxes of 
HD~46223 (the hotter star) and HD~46150 (the cooler star) are also 
displayed. To show that both stars can be modelled by the same mass-loss
rate without affecting the excited frequency domain, we also
plot a 49\,$M_{\sun}$ track computed with 
$\dot{M}=6.8 \times 10^{-8} M_{\sun} {\rm yr}^{-1}$ (magenta-coloured line). 
{\it Right:} Same, but for HD~46966 computed with a mass-loss rate of 
$5.0 \times 10^{-9} M_{\sun} {\rm yr}^{-1}$. 
The full circles indicate positions along the tracks 
that have excited frequencies.}
\label{fig hr_evol}
\end{figure*}

We next apply our standard prewhitening procedure and check the stopping
criteria (Eq.~\ref{eq stop criteria}). From the values found
(Table~\ref{table corot data}), we decide to limit the number
of terms to 500. 
This list of frequencies and corresponding semi-amplitudes and phases is
given in Table~3 (this table is only available in electronic form at CDS).
Applying the significance criterion 
(Eq.~\ref{eq significance}) with the assumption of white noise results
in all 500 terms being significant. The
\citet{1978Ap&SS..56..285L} test shows 296 of these to be unique.
Multifrequency fitting (see Sect.~\ref{sect multifrequency fitting})
gives basically the same set of terms, though with slightly different
amplitudes.

However, it is again clear that there is a substantial red-noise component
present in the spectrum, which we fit with the fit-function
of Eq.~\ref{eq fit function}
(see Table~\ref{table corot data} and Fig.~\ref{fig red noise}, middle).
Judging the significance of peaks with respect to this red noise, we
find 50 significant terms (indicated with 'R' in Table~3),
using a 0.01 cut-off.

When we split
the observed data set into two, we find that 195 of the 500 terms
occur in both halves, but this is less than the simulations of
Sect.~\ref{sect split in two halves} show. Under the red-noise hypothesis,
only seven of the 50 terms are present in both halves. 
We also explore the linear combinations of all 500 frequencies, 
but simulations show these
to be not significant, as for HD~46223 (see Sect.~\ref{sect combinations}).
Among the red-noise frequencies, we find no linear combinations, unless
we relax the criterion we take for considering two frequencies to be equal.
As for HD~46223, this shows that there are no significant linear combinations.

We next apply a similar frequency spacing test as for HD~46223. A peak
in the autocorrelation function is found at $\Delta f$ = 0.58~d$^{-1}$,
but it is not confirmed by the plot with the number of components in a
chain (Fig.~\ref{fig spacing}, middle). Finally, a time-frequency
analysis (Fig.~\ref{fig time-frequency analysis}, middle) shows short
``lifetimes" for the frequencies, which is again consistent with red noise.

\section{HD~46966}
\label{sect HD 46966}

The treatment of the data of HD~46966 is similar to that of
HD~46223 and HD~46150 and details are listed in Table~\ref{table corot data}.
The CoRoT curve of HD~46966 
(Fig.~\ref{fig light curves}, bottom)
exhibits an almost regular
oscillation with an approximate period of 10 days,
but only about three cycles are covered by the observations.
In addition,
we detect variability on shorter timescales. The peak-to-peak
variation amplitudes correspond to about 6 mmag.

Because there is little
power at high frequencies, we limit the frequency domain to 0--100~d$^{-1}$.
We apply our standard prewhitening procedure and check the stopping
criteria (Eq.~\ref{eq stop criteria}). From the values found
(Table~\ref{table corot data}), we decide to limit the number
of terms to 300.
The list of frequencies and corresponding semi-amplitudes and phases is
given in Table~4 (this table is only available in electronic form at CDS). The
\citet{1978Ap&SS..56..285L} criterion shows 196 of these to be unique.
The amplitude spectrum 
(Fig.~\ref{fig spectrum}, bottom)
is characterised by
a first peak that is three times larger than the second one. 
This peak, at
$f$ = 0.084~d$^{-1}$, i.e. 12
days, corresponds to the observed almost regular oscillation.

As for the other two stars, it is better to consider red noise,
described by the fit-function from Eq.~\ref{eq fit function}. 
Applying this, we find 53
significant terms, with a cut-off of 0.01. They are indicated
with an 'R'
in Table~4. Note that in this analysis,
the $f$ = 0.084~d$^{-1}$ frequency discussed above is no longer
considered significant.
Table~4 shows that it has a false-alarm probability of 0.997, 
which is too high
compared to our 0.01 cut-off.
Multifrequency fitting (see Sect.~\ref{sect multifrequency fitting})
gives basically the same set of frequencies, though with slightly different
amplitudes.

The test where we split the observed data set into two, shows that only 
77 of the 300 terms are found in both halves. For the terms
that are significant under the red-noise hypothesis, only three
are found in both halves. Simulations as in
Sect.~\ref{sect split in two halves} show that these numbers are not
significant.

We also explore the linear combinations between the 300 frequencies,
but simulations show that these are not significant, as for the two other
stars. Limiting ourselves to the 53 red-noise frequencies, we find no 
linear combinations. Only when we increase by a factor of three
the criterion we take for equality of frequencies, do we start finding
linear combinations. The result is therefore not significant.

We also apply a similar frequency spacing test as for HD~46223. Again, a peak
in the autocorrelation function is found, at $\Delta f$ = 0.25~d$^{-1}$.
The plot with the number of frequencies in a chain peaks at a
different value of
$\Delta f$ = 0.43~d$^{-1}$, however (Fig.~\ref{fig spacing}, bottom),
showing the result to be not significant.
As for the two other stars, the time-frequency
analysis (Fig.~\ref{fig time-frequency analysis}, bottom) shows short
lifetimes for the frequencies; this again indicates the presence of
red noise.

\section{Theoretical pulsations}
\label{sect asteroseismological modelling}

\begin{figure}
\resizebox{\hsize}{!}{\includegraphics[]{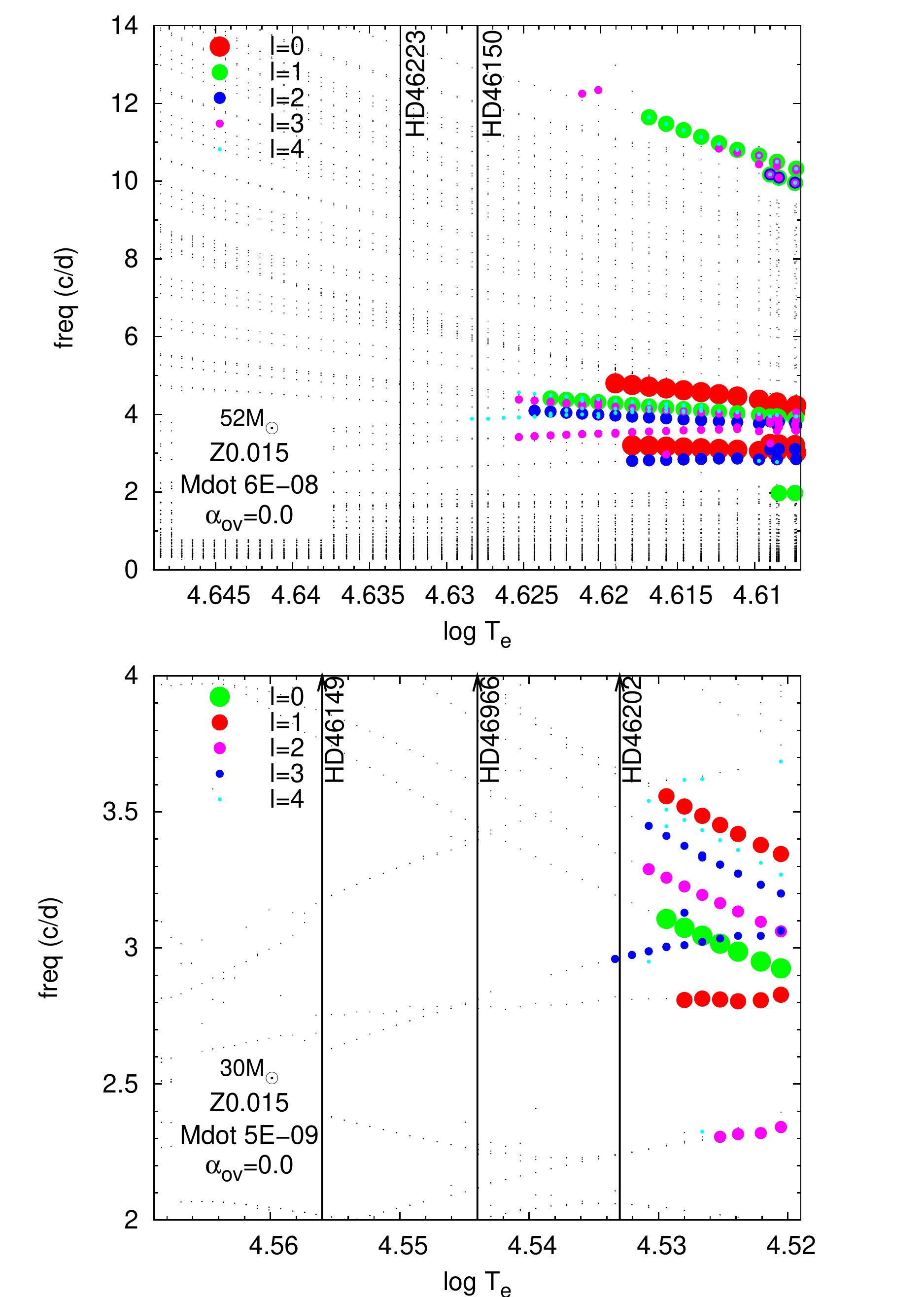}}
\caption{Theoretical oscillation frequencies 
(for the $\kappa$-mechanism) computed with the MAD 
non-adiabatic code, as a function of the effective temperature.
The results shown here do not include overshooting.
{\em Top:} 52\,$M_{\sun}$ track. {\em Bottom:} 30\,$M_{\sun}$
track.
The black points indicate models that 
do not show pulsations, the 
coloured full circles stand for those models with the excited frequencies.
The $l$-value is colour-coded (see legend). 
We also indicate, with vertical lines,
the effective temperatures of the three stars studied
here, as well as HD~46149 \citep{2010A&A...519A..38D} and 
HD~46202 \citep{2011A&A...527A.112B}.}
\label{fig theo_freq}
\end{figure}

An important goal of the CoRoT photometry is to detect radial or non-radial 
pulsation modes in massive stars. To predict which modes can be expected,
we compute a grid of models by 
using the Rome stellar evolution code ATON \citep{2008Ap&SS.316...93V}.
In these calculations, we assume
a metallicity of $Z=0.015$ and the metal mixture of 
\citet{1993pavc.conf..205G}. We also use the 
MAD non-adiabatic 
pulsation code \citep{2003A&A...398..677D} to determine the frequencies of 
excited modes. The computations are made on the basis of 
preliminary stellar and wind 
parameters derived from the atmosphere code CMFGEN
\citep{1998ApJ...496..407H}
by Martins et al. (in prep.). These parameters are listed in
Table~\ref{table stellar parameters}. 
We note however that we use the same value for the mass-loss rate
for HD~46223 as for HD~46150 to compute the evolutionary tracks.
The difference between the value used and those listed
in Table~\ref{table stellar parameters} does not significantly affect the
results mentioned in the present paper.
Additionally, we consider both models with and without overshooting.

\addtocounter{table}{2} 
\begin{table}
\caption{Stellar parameters for the three stars.}
\label{table stellar parameters}
\centering
\begin{tabular}{lccc}
\hline\hline
Parameter & HD\,46223 & HD\,46150 & HD\,46966\\
\hline
$\log T_{\rm{eff}}{\rm (K)}$ & $4.633 \pm 0.020$ & $4.628 \pm 0.021$ &  $4.544 \pm 0.025$ \\
$\log g$ (cgs)      & $4.00$ & $4.00$ & $3.75$ \\
$\log L/L_{\sun}$       & $5.60 \pm 0.11$ & $5.65 \pm 0.11$ & $5.15 \pm 0.22$\\
$R/R_{\sun}$            & 11.39 & 12.35 & 10.24\\
$\log \dot{M}~(M_{\sun} {\rm yr}^{-1})$   & $-7.20$ & $-7.17$ & $-8.30$ \\
$v_{\rm eq} \sin\,i $ (km s$^{-1}$) &100 &100 &50 \\
$v_{\rm macro}$ (km s$^{-1}$) &32& 37& 27\\
\hline
\end{tabular}
\tablefoot{Preliminary stellar and wind parameters derived by Martins et al. (in prep.).}
\end{table}

HD~46223 and HD~46150 are in very similar positions in the
HR diagram, whilst 
HD~46966 is cooler and more evolved. All three 
stars lie just at the boundary of the instability strip of the 
$\kappa$-mechanism. Theoretical excited frequencies 
for that mechanism are generally
in the range of 2--5~d$^{-1}$ for $l=0$ to $l=4$ and 
a non-radial mode is also excited around 12~d$^{-1}$. 
The theoretical frequencies from our model calculations
are shown on Fig.~\ref{fig theo_freq} for 30 and 52\,$M_{\sun}$
(without overshooting): 
models with these masses 
present the largest number of excited modes. 
Models computed with overshooting give similar results.

For HD~46223, we consider stellar evolution models with an initial
mass ranging from 
45 to 53\,$M_{\sun}$ (Fig.~\ref{fig hr_evol}, left panel). On these tracks,
models with a mass greater than 47\,$M_{\sun}$ show excited modes. 
For this mass
range, the excited modes all have frequencies around 4~d$^{-1}$.
However, if we look at the observed amplitude spectrum of
HD~46223 (Fig.~\ref{fig spectrum}, top),
no outstanding peak is situated close to 4~d$^{-1}$ which means that this star 
probably does not present pulsations, neither radial nor 
non-radial. 
For HD~46150, a similar conclusion can be drawn as for HD~46223. 
There are low-amplitude peaks around 4~d$^{-1}$, but 
the peaks are not strong and they could be the result of noise.

The amplitude spectrum of HD~46966 is clearly dominated by 
peaks at
the frequencies $f=0.084$ and $0.039$~d$^{-1}$. 
The time coverage of the CoRoT observations is too short to know
if these are regular variations. Furthermore, it is unlikely that
these two frequencies are due to
excited modes (Fig.~\ref{fig theo_freq}, 
bottom) since those modes are predicted to be between 
2.5 and 4~d$^{-1}$. Even though the amplitude spectrum has 
peaks
situated in this frequency region, these peaks are not strong.

It is important to emphasize that the results presented in this section
are dependent on the stellar parameters determined by 
Martins et al. (in prep.). These parameters are still under investigation and
should therefore be considered as preliminary. Nonetheless, the present
work shows that even with the unprecedented quality of the CoRoT light 
curves, 
we do not convincingly detect the predicted excited modes. This
suggests they are only weakly excited, or perhaps, not at all.
On the other hand, a number of stronger peaks are observed, but at
frequencies that are so different from the theoretical ones that we
cannot attribute them to the expected excited modes, at least not within
the framework of current theoretical models.
It is thus not possible to conclude with certainty that these 
three stars are pulsating. 
Moreover, we also explored the possibility of solar-like oscillations, since
\citet{2010A&A...519A..38D} found such oscillations in the OB binary
system HD~46149. But 
the existence of such oscillations was rejected in HD~46150,
HD~46223 and HD~46966 because no regular spacings were found in the 
amplitude spectra.

\section{Discussion}
\label{sect discussion}

\subsection{Radius and rotational frequency}
\label{sect radius and rotational frequency}

In a presumably single star, large-scale photometric variations can have a 
frequency compatible with the rotational frequency of the star. Their exact 
origin could be attributed to the existence of a spot on the star or be
associated with the stellar wind.

The first low frequency detected in the amplitude spectrum of HD~46223 
is $f = 0.249$~d$^{-1}$, corresponding to 4.02~days. With this period and the 
parameters from Table~\ref{table stellar parameters},
we derive an equatorial velocity $v_{\mathrm{eq}}$ 
of about 140~km\,s$^{-1}$ which is greater than the $v_{\mathrm{eq}} \sin i$ 
value of 100~km\,s$^{-1}$ measured by \citet{2009A&A...502..937M} and
Martins et al.~(in prep.). Consequently, the frequency $f = 0.249$~d$^{-1}$ 
could be associated with the rotational cycle of the star, implying a rotational 
period of about $\sim\,4$~days and an inclination of $43-45^{\circ}$.

For HD~46150, the amplitude spectrum exhibits a strong peak at 
$f = 0.055$~d$^{-1}$, corresponding to 18~days. The derived equatorial 
velocity is 34~km\,s$^{-1}$, which is too low compared to the 
$v_{\mathrm{eq}} \sin i$. The frequency corresponding to the equatorial 
velocity found by \cite{2009A&A...502..937M} or
Martins et al.,
should be close to or larger than 0.16~d$^{-1}$. 
The list of frequencies (Table~3) shows that the 
third frequency ($f=0.144$~d$^{-1}$) in Table~3 is quite close.
It corresponds to an equatorial velocity of 90~km\,s$^{-1}$, which is within
acceptable range of the \citeauthor{2009A&A...502..937M} and Martins et al.
value, but would require an inclination of $90^{\circ}$.

The amplitude spectrum of HD~46966 is marked by a strong peak at 
$f = 0.084$~d$^{-1}$, representing a period of 12~days. With the stellar 
parameters mentioned in 
Table~\ref{table stellar parameters}, 
we estimate the equatorial velocity to be 
close to 44~km\,s$^{-1}$. 
This velocity is just below the 1-sigma error bar on the 
$v_{\mathrm{eq}} \sin i$ determined by \cite{2009A&A...502..937M}.
Martins et al.~(in prep.), on the other hand,
clearly show 
that the $v_{\mathrm{eq}} \sin i$ is smaller than 50~km\,s$^{-1}$. 
Consequently, the frequency $f = 0.084$~d$^{-1}$ could be associated with the 
rotational cycle of the star, implying a rotational period of about 
12~days.

\subsection{Pulsations and red noise}
\label{sect pulsations and red noise}

For the three O-type stars studied here, a substantial number of
sine function terms are found. However, it is unclear how many of those
-- if any -- can be classified as true independent 
modes.
They do not show the linear combinations
or frequency spacings that may be expected in pulsators. 
If these terms were interpreted as pulsational modes, then their 
lifetimes would be very short.
The amplitude spectra 
for these three stars 
are therefore unlike any of those of known pulsators
in that part of the HR diagram \citep[][Chapter 2]{2010aste.book.....A}.

This is clearly illustrated by comparing the periodogram of 
these hot O stars (spectral types O4 -- O8)
to that of the cooler star HD~46202 (O9 V).
This star behaves like a classical pulsator, with clear $\beta$ Cep-like
pulsations that dominate the power spectrum 
\citep{2011A&A...527A.112B}. 
Various causes for what we see in our periodograms can be eliminated:
rotational effects on pulsation can be excluded as 
our periodograms are very different from those of Be
stars \citep[e.g.][]{2009A&A...506..125D}. Spots on the surface of
the star would give clearly isolated frequencies (and linear combinations
between them). A power excess due to stochastic modes would occur at
higher frequencies.

The interpretation that ascribes most of the power in the periodogram 
to red noise therefore seems more appropriate. 
We stress again that 
this red noise is caused by a physical mechanism in the star. It is not
an instrumental effect, because its behaviour is different from
star to star in these simultaneous observations.
The parameters of the fit-function (Table~\ref{table corot data})
show that 
HD~46966 has a somewhat lower $\alpha_0$ and a somewhat longer
$\tau$ than the other two stars. Compared with Plaskett's
Star (HD~47129) a clearer difference is seen:
especially the $\gamma$ value is different
\citep[$\gamma$ = 2.3,][]{2011A&A...525A.101M}.
We also fitted the red-noise component of HD~46149 \citep{2010A&A...519A..38D}.
This star has a $\tau = 29.5 \pm 12$ and 
$\gamma=0.55 \pm 0.01$, which are substantially
different from the stars studied in this paper. All this points to
an origin for the red noise that is intrinsic to the star.

To see how convincing the red-noise description is, we simulate
light curves containing only red noise (according to the 
fit-function -- Eq.~\ref{eq fit function}), and analyse them in the same way
as the real data. We follow the prescription by \citet{1995A&A...300..707T}
to generate the red noise. In the simulated data, the number of 
significant red-noise
frequencies found is somewhat higher (by a factor less than two) 
than in the real
data. Only one or two of these significant red-noise frequencies can be
found in both halves of the light curve (somewhat less than for the real
data). The number of linear combinations
found is higher than for the real data, but this is presumably related
to the higher number of significant red-noise frequencies.
It therefore seems that the observed data can largely be described
as red noise, though a few pulsational frequencies might be present.

The presence of red noise in the stellar signal 
indicates stochastic, chaotic or quasiperiodic effects. It occurs
in a number of astrophysical contexts, such as 
the light curves of Mira variables \citep{2009ApJ...691.1470T}
and red supergiants \citep{2006MNRAS.372.1721K},
and the X-ray light curves of active galaxies \citep{2003MNRAS.345.1271V},
dwarf novae \citep{2004A&A...420..273H}
and high-mass X-ray binaries \citep{1994ASPC...61..353B}.

The physical cause of this red noise in early-type
stars is not clear.
We next 
discuss three possibilities: sub-surface convection, granulation and
inhomogeneities in the stellar wind. We stress that all three are highly
speculative, and that as yet no modelling exists to see if they would
give the correct quantitative behaviour.

A link between red noise and convection was already pointed out by 
\citet{1975ApJ...195..137S}. For early-type stars, \citet{2009A&A...499..279C}
show the existence of a sub-surface convection zone, caused by the
opacity peak due to iron ionization. They show that stars in the 
HR diagram where this
zone is present also have microturbulence, non-radial pulsations,
wind clumping and line-profile variability, suggesting that these are
caused by the convection zone. We speculate that the red noise detected
in the early-type stars studied here is also a consequence of this
convection zone, and could be associated with any of the above
observational indicators. The position of our stars in the HR diagram
shows that they should have a rather weak convection zone
\citep[see][their Fig.~9]{2009A&A...499..279C}. This suggests that 
the high quality of the CoRoT data 
allows us to see a more subtle effect
than \citeauthor{2009A&A...499..279C} considered.
Instead of attributing the red noise to convection only, there could also
be an 
interaction between pulsation and convection. \citet{2009EAS....38...43P}
shows that this can result in 
red-noise dominated light curves.

We also explore if the physical cause of the red noise could be described 
as granulation.
\citet{2010ApJ...711L..35K} show that, in the two $\delta$~Scuti
stars they study, the many hundreds of peaks in the amplitude
spectrum can also be interpreted as being due to granulation. Their Fig.~3
relates the typical frequency of granulation (i.e. the inverse of the
timescale) to a specific combination
of mass, radius and effective temperature. Although this relation
is based on cooler stars, we nevertheless use it to predict the
granulation frequency in our stars. The value we find is $\sim~35$
d$^{-1}$, but this cannot be linked to any significant frequency or inverse
timescale of the observations discussed in this paper. 
If granulation is the cause of the red noise in these hot O stars, 
it would therefore
need to have properties that cannot simply be extrapolated from
granulation in cooler type stars.

Finally, the red noise could be due to inhomogeneities in the stellar wind.
There are various observational indicators showing that the stellar winds
of early-type stars are clumped 
\citep[see review by][and references therein]{2008A&ARv..16..209P}.
The exact cause of this clumping is not clear. Intrinsic instabilities
in the radiative driving mechanism have been proposed
\citep{1984ApJ...284..337O}, but the onset of clumping seems to
happen very close to the stellar surface \citep{2006A&A...454..625P}.
The red noise could therefore be related to the onset of clumping at
the stellar surface.

The behaviour of the early-type O stars (spectral types O4 -- O8) 
is considerably different from that of later-type stars. 
When taking out the dominant binary signature of HD~47129
(O8 III/I + O7.5 V/III), a series of frequencies is found that
might be due to pulsations, but a clear red-noise component is
present as well
\citep{2011A&A...525A.101M}.
HD~46149 \citep[O8, ][]{2010A&A...519A..38D}
does have significant frequency spacing.
%
HD~46202 (O9 V) shows clear $\beta$ Cep-like pulsations
\citep{2011A&A...527A.112B}.
Observations with the {\em MOST (Microvariability and Oscillations
of Stars)} satellite of the O9.5 V star $\zeta$~Oph 
also show $\beta$ Cep-type pulsations \citep{2005ApJ...623L.145W}.
Among the early B-type stars, pulsation frequencies are 
easily found:
\citet{2009A&A...506..471D} present
a study of 358 candidate B-star pulsators, showing numerous classical
SPBs (Slowly-Pulsating B stars).
We also note that a red-noise component is present
in their spectra, but it is much more limited in frequency
range than for the stars discussed here (see their Fig.~2).


The present data, combined with information from the literature, permit
a first attempt at empirically mapping the hottest part of the instability
strip for $\beta$ Cep and SPB stars. We can compare
the position in the HR diagram of our stars,
HD~47129 \citep{2011A&A...525A.101M},
HD~46149 \citep{2010A&A...519A..38D},
HD~46202 \citep{2011A&A...527A.112B} and
$\zeta$~Oph \citep{2005ApJ...623L.145W} 
to theoretical predictions of the $\beta$ Cep and SPB instability strip.
\citet{2008JPhCS.118a2079Z} present the theoretical strip for
models up to 40\,$M_{\sun}$ for a metallicity $Z=0.02$.
Most of the stars in our list fall into their theoretical $\beta$ Cep 
instability strip, but outside the SPB one. As only a few of these stars
actually
show $\beta$ Cep pulsations, this suggests either that the strip is much
narrower, or that the pulsations are excited for only a small fraction
of the stars in the strip.
Especially constraining is a group of three stars (HD~46966, HD~46202
and $\zeta$~Oph) which are close to one another in the HR diagram,
with HD~46202 and $\zeta$~Oph showing $\beta$ Cep pulsations, but
HD~46966 not.

A number of O-type $\beta$ Cep pulsators have also been detected in 
ground-based observations
\citep[e.g.][]{2006A&A...452..945T, 2007A&A...463..243D, 2008A&A...477..917P}.
These are all late O-type stars and therefore close to the
three-star group. A detailed knowledge of their astrophysical
parameters would allow a better mapping of the instability strip
in that part of the HR diagram. In the hotter part, we note the absence
of $\beta$ Cep pulsations in the main-sequence stars discussed here.
The theoretical $\beta$ Cep strip extends to the terminal age main sequence 
(TAMS) however, and stars more evolved than those studied here
could therefore show
$\beta$ Cep pulsations.

\section{Conclusions}
\label{sect conclusions}

The CoRoT light curves of the hot O-type stars HD~46223, HD~46150 and
HD~46966 were analysed using standard methods to search for 
pulsation frequencies. The detection of such frequencies would allow
an asteroseismological interpretation. However, the results show that most
of the variations of these stars are of a stochastic nature. 
The only significant exception to the above is the possible rotation period
detected for HD~46223 and HD~46966, although these conclusions are not 
strong and still require confirmation. Evidence for a rotation period
in the HD~46150 data is even less strong.

The periodogram of these three stars is clearly different from that
of a classical pulsator such as HD~46202.
Most of the power in our periodograms is not due to pulsation, but
is more correctly described by red noise. 
Based on the limited number of stars studied so far,
a trend is suggested: the earliest O stars show red noise, while the
later O-types have pulsational frequencies of the $\beta$ Cep type.
The switch-over occurs around spectral type O8. 

The specific physical cause of this noise in O-type stars is at present
unclear. 
We point out the possibilities of sub-surface convection, granulation
or inhomogeneities in the wind. All three options are highly speculative
and therefore await confirmation by detailed modelling.
The CoRoT data of these three hot stars
present interesting challenges for our understanding of the outer
photosphere and inner wind of early-type stars.

\begin{acknowledgements}
We thank the CoRoT team for the acquisition and
the reduction of the CoRoT data. 
We also thank F. Martins for making the preliminary
results of his analysis available to us.
We thank the anonymous referee for his/her constructive comments.
L.M, E.G. and G.R. are grateful to the F.N.R.S
(Belgium), the PRODEX XMM/Integral contract (Belspo), GAIA DPAC Prodex
and the Communaut\'e fran\c{c}aise de Belgique -- 
Action de recherche concert\'ee -- A.R.C. -- 
Acad\'emie Wallonie-Europe for their support. 
P.D. and C.A. acknowledge
the financial support of the European Research Council under the
European Community's Seventh Framework Programme (FP7/2007--2013)/ERC
grant agreement n$^{\rm o}$ 227224 (PROSPERITY), from the Research Council of
K.U.Leuven (GOA/2008/04), and from the Belgian federal science policy office
(C90309: CoRoT Data Exploitation). 
\end{acknowledgements}

\bibliographystyle{aa}
\bibliography{Ostars}

\begin{thebibliography}{63}
\expandafter\ifx\csname natexlab\endcsname\relax\def\natexlab#1{#1}\fi

\bibitem[{{Aerts} {et~al.}(2010){Aerts}, {Christensen-Dalsgaard}, \&
  {Kurtz}}]{2010aste.book.....A}
{Aerts}, C., {Christensen-Dalsgaard}, J., \& {Kurtz}, D.~W. 2010,
  {Asteroseismology} ({Springer, Dordrecht})

\bibitem[{{Auvergne} {et~al.}(2009){Auvergne}, {Bodin}, {Boisnard}, {Buey},
  {Chaintreuil}, {Epstein}, {Jouret}, {Lam-Trong}, {Levacher}, {Magnan},
  {Perez}, {Plasson}, {Plesseria}, {Peter}, {Steller}, {Tiph{\`e}ne}, {Baglin},
  {Agogu{\'e}}, {Appourchaux}, {Barbet}, {Beaufort}, {Bellenger}, {Berlin},
  {Bernardi}, {Blouin}, {Boumier}, {Bonneau}, {Briet}, {Butler}, {Cautain},
  {Chiavassa}, {Costes}, {Cuvilho}, {Cunha-Parro}, {de Oliveira Fialho},
  {Decaudin}, {Defise}, {Djalal}, {Docclo}, {Drummond}, {Dupuis}, {Exil},
  {Faur{\'e}}, {Gaboriaud}, {Gamet}, {Gavalda}, {Grolleau}, {Gueguen},
  {Guivarc'h}, {Guterman}, {Hasiba}, {Huntzinger}, {Hustaix}, {Imbert},
  {Jeanville}, {Johlander}, {Jorda}, {Journoud}, {Karioty}, {Kerjean},
  {Lafond}, {Lapeyrere}, {Landiech}, {Larqu{\'e}}, {Laudet}, {Le Merrer},
  {Leporati}, {Leruyet}, {Levieuge}, {Llebaria}, {Martin}, {Mazy}, {Mesnager},
  {Michel}, {Moalic}, {Monjoin}, {Naudet}, {Neukirchner}, {Nguyen-Kim},
  {Ollivier}, {Orcesi}, {Ottacher}, {Oulali}, {Parisot}, {Perruchot},
  {Piacentino}, {Pinheiro da Silva}, {Platzer}, {Pontet}, {Pradines},
  {Quentin}, {Rohbeck}, {Rolland}, {Rollenhagen}, {Romagnan}, {Russ}, {Samadi},
  {Schmidt}, {Schwartz}, {Sebbag}, {Smit}, {Sunter}, {Tello}, {Toulouse},
  {Ulmer}, {Vandermarcq}, {Vergnault}, {Wallner}, {Waultier}, \&
  {Zanatta}}]{2009A&A...506..411A}
{Auvergne}, M., {Bodin}, P., {Boisnard}, L., {et~al.} 2009, \aap, 506, 411

\bibitem[{{Baglin} {et~al.}(2006){Baglin}, {Auvergne}, {Barge}, {Deleuil},
  {Catala}, {Michel}, {Weiss}, \& {The COROT Team}}]{2006ESASP1306...33B}
{Baglin}, A., {Auvergne}, M., {Barge}, P., {et~al.} 2006, in ESA Special
  Publication, Vol. 1306, The CoRoT Mission Pre-Launch Status - Stellar
  Seismology and Planet Finding, ed. {M.~Fridlund, A.~Baglin, J.~Lochard, \&
  L.~Conroy} ({ESA, Noordwijk}), 33

\bibitem[{{Bisiacchi} {et~al.}(1982){Bisiacchi}, {Lopez}, \&
  {Firmani}}]{1982A&A...107..252B}
{Bisiacchi}, G.~F., {Lopez}, J.~A., \& {Firmani}, C. 1982, \aap, 107, 252

\bibitem[{{Bonatto} \& {Bica}(2009)}]{2009MNRAS.394.2127B}
{Bonatto}, C. \& {Bica}, E. 2009, \mnras, 394, 2127

\bibitem[{{Briquet} {et~al.}(2011){Briquet}, {Aerts}, {Baglin}, {Nieva},
  {Degroote}, {Przybilla}, {Noels}, {Schiller}, {Vu{\v c}kovi{\'c}}, {Oreiro},
  {Smolders}, {Auvergne}, {Baudin}, {Catala}, {Michel}, \&
  {Samadi}}]{2011A&A...527A.112B}
{Briquet}, M., {Aerts}, C., {Baglin}, A., {et~al.} 2011, \aap, 527, A112

\bibitem[{{Burderi}(1994)}]{1994ASPC...61..353B}
{Burderi}, L. 1994, in Astronomical Society of the Pacific Conference Series,
  Vol.~61, Astronomical Data Analysis Software and Systems III, ed.
  {D.~R.~Crabtree, R.~J.~Hanisch, \& J.~Barnes} ({ASP, California}), 353

\bibitem[{{Cantiello} {et~al.}(2009){Cantiello}, {Langer}, {Brott}, {de Koter},
  {Shore}, {Vink}, {Voegler}, {Lennon}, \& {Yoon}}]{2009A&A...499..279C}
{Cantiello}, M., {Langer}, N., {Brott}, I., {et~al.} 2009, \aap, 499, 279

\bibitem[{{Conti} \& {Ebbets}(1977)}]{1977ApJ...213..438C}
{Conti}, P.~S. \& {Ebbets}, D. 1977, \apj, 213, 438

\bibitem[{{Conti} \& {Leep}(1974)}]{1974ApJ...193..113C}
{Conti}, P.~S. \& {Leep}, E.~M. 1974, \apj, 193, 113

\bibitem[{{Conti} {et~al.}(1977){Conti}, {Leep}, \&
  {Lorre}}]{1977ApJ...214..759C}
{Conti}, P.~S., {Leep}, E.~M., \& {Lorre}, J.~J. 1977, \apj, 214, 759

\bibitem[{{De Cat} {et~al.}(2007){De Cat}, {Briquet}, {Aerts}, {Goossens},
  {Saesen}, {Cuypers}, {Yakut}, {Scuflaire}, {Dupret}, {Uytterhoeven}, {van
  Winckel}, {Raskin}, {Davignon}, {Le Guillou}, {van Malderen}, {Reyniers},
  {Acke}, {De Meester}, {Vanautgaerden}, {Vandenbussche}, {Verhoelst},
  {Waelkens}, {Deroo}, {Reyniers}, {Ausseloos}, {Broeders},
  {Daszy{\'n}ska-Daszkiewicz}, {Debosscher}, {De Ruyter}, {Lefever}, {Decin},
  {Kolenberg}, {Mazumdar}, {van Kerckhoven}, {De Ridder}, {Drummond}, {Barban},
  {Vanhollebeke}, {Maas}, \& {Decin}}]{2007A&A...463..243D}
{De Cat}, P., {Briquet}, M., {Aerts}, C., {et~al.} 2007, \aap, 463, 243

\bibitem[{{Degroote} {et~al.}(2009{\natexlab{a}}){Degroote}, {Aerts},
  {Ollivier}, {Miglio}, {Debosscher}, {Cuypers}, {Briquet}, {Montalb{\'a}n},
  {Thoul}, {Noels}, {De Cat}, {Balaguer-N{\'u}{\~n}ez}, {Maceroni}, {Ribas},
  {Auvergne}, {Baglin}, {Deleuil}, {Weiss}, {Jorda}, {Baudin}, \&
  {Samadi}}]{2009A&A...506..471D}
{Degroote}, P., {Aerts}, C., {Ollivier}, M., {et~al.} 2009{\natexlab{a}}, \aap,
  506, 471

\bibitem[{{Degroote} {et~al.}(2010){Degroote}, {Briquet}, {Auvergne},
  {Sim{\'o}n-D{\'{\i}}az}, {Aerts}, {Noels}, {Rainer}, {Hareter}, {Poretti},
  {Mahy}, {Oreiro}, {Vu{\v c}kovi{\'c}}, {Smolders}, {Baglin}, {Baudin},
  {Catala}, {Michel}, \& {Samadi}}]{2010A&A...519A..38D}
{Degroote}, P., {Briquet}, M., {Auvergne}, M., {et~al.} 2010, \aap, 519, A38

\bibitem[{{Degroote} {et~al.}(2009{\natexlab{b}}){Degroote}, {Briquet},
  {Catala}, {Uytterhoeven}, {Lefever}, {Morel}, {Aerts}, {Carrier}, {Auvergne},
  {Baglin}, \& {Michel}}]{2009A&A...506..111D}
{Degroote}, P., {Briquet}, M., {Catala}, C., {et~al.} 2009{\natexlab{b}}, \aap,
  506, 111

\bibitem[{{Diago} {et~al.}(2009){Diago}, {Guti{\'e}rrez-Soto}, {Auvergne},
  {Fabregat}, {Hubert}, {Floquet}, {Fr{\'e}mat}, {Garrido}, {Andrade}, {de
  Batz}, {Emilio}, {Espinosa Lara}, {Huat}, {Janot-Pacheco}, {Leroy},
  {Martayan}, {Neiner}, {Semaan}, {Suso}, {Catala}, {Poretti}, {Rainer},
  {Uytterhoeven}, {Michel}, \& {Samadi}}]{2009A&A...506..125D}
{Diago}, P.~D., {Guti{\'e}rrez-Soto}, J., {Auvergne}, M., {et~al.} 2009, \aap,
  506, 125

\bibitem[{{Dupret} {et~al.}(2003){Dupret}, {De Ridder}, {De Cat}, {Aerts},
  {Scuflaire}, {Noels}, \& {Thoul}}]{2003A&A...398..677D}
{Dupret}, M.-A., {De Ridder}, J., {De Cat}, P., {et~al.} 2003, \aap, 398, 677

\bibitem[{{Garmany} {et~al.}(1980){Garmany}, {Conti}, \&
  {Massey}}]{1980ApJ...242.1063G}
{Garmany}, C.~D., {Conti}, P.~S., \& {Massey}, P. 1980, \apj, 242, 1063

\bibitem[{{Gosset}(2007)}]{Gosset2007}
{Gosset}, E. 2007, {Le calcul du niveau de signification du plus haut pic dans
  les p\'eriodogrammes de type Fourier par la formule de Horne-Baliunas est
  contreindiqu\'e, Th\`ese d'Agr\'egation de l'Enseignement Sup\'erieur,
  Seconde Th\`ese Annexe (University of Li\`ege, Li\`ege)}

\bibitem[{{Gosset} {et~al.}(2001){Gosset}, {Royer}, {Rauw}, {Manfroid}, \&
  {Vreux}}]{2001MNRAS.327..435G}
{Gosset}, E., {Royer}, P., {Rauw}, G., {Manfroid}, J., \& {Vreux}, J. 2001,
  \mnras, 327, 435

\bibitem[{{Grevesse} \& {Noels}(1993)}]{1993pavc.conf..205G}
{Grevesse}, N. \& {Noels}, A. 1993, in Perfectionnement de l'Association
  Vaudoise des Chercheurs en Physique (AVCP, Lausanne), 205

\bibitem[{{Hakala} {et~al.}(2004){Hakala}, {Ramsay}, {Wheatley}, {Harlaftis},
  \& {Papadimitriou}}]{2004A&A...420..273H}
{Hakala}, P., {Ramsay}, G., {Wheatley}, P., {Harlaftis}, E.~T., \&
  {Papadimitriou}, C. 2004, \aap, 420, 273

\bibitem[{{Hannan}(1980)}]{HANNAN}
{Hannan}, E.~J. 1980, Annals of Statistics, 8, 1071

\bibitem[{{Hannan} \& {Quinn}(1979)}]{HQC}
{Hannan}, E.~J. \& {Quinn}, B. 1979, Journal of the Royal Statistical Society,
  Series B, 41, 190

\bibitem[{{Heck} {et~al.}(1985){Heck}, {Manfroid}, \&
  {Mersch}}]{1985A&AS...59...63H}
{Heck}, A., {Manfroid}, J., \& {Mersch}, G. 1985, \aaps, 59, 63, (HMM)

\bibitem[{{Hensberge} {et~al.}(2000){Hensberge}, {Pavlovski}, \&
  {Verschueren}}]{2000A&A...358..553H}
{Hensberge}, H., {Pavlovski}, K., \& {Verschueren}, W. 2000, \aap, 358, 553

\bibitem[{{Hillier} \& {Miller}(1998)}]{1998ApJ...496..407H}
{Hillier}, D.~J. \& {Miller}, D.~L. 1998, \apj, 496, 407

\bibitem[{{Hiltner}(1956)}]{1956ApJS....2..389H}
{Hiltner}, W.~A. 1956, \apjs, 2, 389

\bibitem[{{Howarth} \& {Reid}(1993)}]{1993A&A...279..148H}
{Howarth}, I.~D. \& {Reid}, A.~H.~N. 1993, \aap, 279, 148

\bibitem[{{Huat} {et~al.}(2009){Huat}, {Hubert}, {Baudin}, {Floquet}, {Neiner},
  {Fr{\'e}mat}, {Guti{\'e}rrez-Soto}, {Andrade}, {de Batz}, {Diago}, {Emilio},
  {Espinosa Lara}, {Fabregat}, {Janot-Pacheco}, {Leroy}, {Martayan}, {Semaan},
  {Suso}, {Auvergne}, {Catala}, {Michel}, \& {Samadi}}]{2009A&A...506...95H}
{Huat}, A.-L., {Hubert}, A.-M., {Baudin}, F., {et~al.} 2009, \aap, 506, 95

\bibitem[{{Kallinger} \& {Matthews}(2010)}]{2010ApJ...711L..35K}
{Kallinger}, T. \& {Matthews}, J.~M. 2010, \apjl, 711, L35

\bibitem[{{Kambe} {et~al.}(1997){Kambe}, {Hirata}, {Ando}, {Cuypers}, {Katoh},
  {Kennelly}, {Walker}, {Stefl}, \& {Tarasov}}]{1997ApJ...481..406K}
{Kambe}, E., {Hirata}, R., {Ando}, H., {et~al.} 1997, \apj, 481, 406

\bibitem[{{Kiss} {et~al.}(2006){Kiss}, {Szab{\'o}}, \&
  {Bedding}}]{2006MNRAS.372.1721K}
{Kiss}, L.~L., {Szab{\'o}}, G.~M., \& {Bedding}, T.~R. 2006, \mnras, 372, 1721

\bibitem[{{Liddle}(2007)}]{2007MNRAS.377L..74L}
{Liddle}, A.~R. 2007, \mnras, 377, L74

\bibitem[{{Lomb}(1976)}]{1976Ap&SS..39..447L}
{Lomb}, N.~R. 1976, \apss, 39, 447

\bibitem[{{Loumos} \& {Deeming}(1978)}]{1978Ap&SS..56..285L}
{Loumos}, G.~L. \& {Deeming}, T.~J. 1978, \apss, 56, 285

\bibitem[{{Lucy} \& {Sweeney}(1971)}]{1971AJ.....76..544L}
{Lucy}, L.~B. \& {Sweeney}, M.~A. 1971, \aj, 76, 544

\bibitem[{{Mahy} {et~al.}(2011){Mahy}, {Gosset}, {Baudin}, {Rauw}, {Godart},
  {Morel}, {Degroote}, {Aerts}, {Blomme}, {Cuypers}, {Noels}, {Michel},
  {Baglin}, {Auvergne}, {Catala}, \& {Samadi}}]{2011A&A...525A.101M}
{Mahy}, L., {Gosset}, E., {Baudin}, F., {et~al.} 2011, \aap, 525, A101

\bibitem[{{Mahy} {et~al.}(2009){Mahy}, {Naz{\'e}}, {Rauw}, {Gosset}, {De
  Becker}, {Sana}, \& {Eenens}}]{2009A&A...502..937M}
{Mahy}, L., {Naz{\'e}}, Y., {Rauw}, G., {et~al.} 2009, \aap, 502, 937

\bibitem[{{Ma{\'{\i}}z-Apell{\'a}niz}
  {et~al.}(2004){Ma{\'{\i}}z-Apell{\'a}niz}, {Walborn}, {Galu{\'e}}, \&
  {Wei}}]{2004ApJS..151..103M}
{Ma{\'{\i}}z-Apell{\'a}niz}, J., {Walborn}, N.~R., {Galu{\'e}}, H.~{\'A}., \&
  {Wei}, L.~H. 2004, \apjs, 151, 103

\bibitem[{{Massey} {et~al.}(1995){Massey}, {Johnson}, \&
  {Degioia-Eastwood}}]{1995ApJ...454..151M}
{Massey}, P., {Johnson}, K.~E., \& {Degioia-Eastwood}, K. 1995, \apj, 454, 151

\bibitem[{{Montgomery} \& {O'Donoghue}(1999)}]{1999DSSN...13...28M}
{Montgomery}, M.~H. \& {O'Donoghue}, D. 1999, Delta Scuti Star Newsletter, 13,
  28

\bibitem[{{Munari} \& {Tomasella}(1999)}]{1999A&AS..137..521M}
{Munari}, U. \& {Tomasella}, L. 1999, \aaps, 137, 521

\bibitem[{{Owocki} \& {Rybicki}(1984)}]{1984ApJ...284..337O}
{Owocki}, S.~P. \& {Rybicki}, G.~B. 1984, \apj, 284, 337

\bibitem[{{Perdang}(2009)}]{2009EAS....38...43P}
{Perdang}, J. 2009, in EAS Publications Series, Vol.~38, EAS Publications
  Series, ed. {M.~Goupil, Z.~Kol{\'a}th, N.~Nardetto, \& P.~Kervella} ({EDP,
  Les Ulis, France}), 43

\bibitem[{{Pigulski} \& {Pojma{\'n}ski}(2008)}]{2008A&A...477..917P}
{Pigulski}, A. \& {Pojma{\'n}ski}, G. 2008, \aap, 477, 917

\bibitem[{{Press} {et~al.}(1992){Press}, {Teukolsky}, {Vetterling}, \&
  {Flannery}}]{1992nrfa.book.....P}
{Press}, W.~H., {Teukolsky}, S.~A., {Vetterling}, W.~T., \& {Flannery}, B.~P.
  1992, {Numerical recipes in FORTRAN. The art of scientific computing}
  (University Press, Cambridge, 2nd ed.)

\bibitem[{{Puls} {et~al.}(2006){Puls}, {Markova}, {Scuderi}, {Stanghellini},
  {Taranova}, {Burnley}, \& {Howarth}}]{2006A&A...454..625P}
{Puls}, J., {Markova}, N., {Scuderi}, S., {et~al.} 2006, \aap, 454, 625

\bibitem[{{Puls} {et~al.}(2008){Puls}, {Vink}, \&
  {Najarro}}]{2008A&ARv..16..209P}
{Puls}, J., {Vink}, J.~S., \& {Najarro}, F. 2008, \aapr, 16, 209

\bibitem[{{Rauw} {et~al.}(2008){Rauw}, {De Becker}, {van Winckel}, {Aerts},
  {Eenens}, {Lefever}, {Vandenbussche}, {Linder}, {Naz{\'e}}, \&
  {Gosset}}]{2008A&A...487..659R}
{Rauw}, G., {De Becker}, M., {van Winckel}, H., {et~al.} 2008, \aap, 487, 659

\bibitem[{{Samadi} {et~al.}(2007){Samadi}, {Fialho}, {Costa}, {Drummond},
  {Pinheiro Da Silva}, {Baudin}, {Boumier}, \& {Jorda}}]{2007astro.ph..3354S}
{Samadi}, R., {Fialho}, F., {Costa}, J.~E.~S., {et~al.} 2007, ArXiv
  Astrophysics e-prints, astro-ph/0703354

\bibitem[{{Scargle}(1982)}]{1982ApJ...263..835S}
{Scargle}, J.~D. 1982, \apj, 263, 835

\bibitem[{{Schwarzschild}(1975)}]{1975ApJ...195..137S}
{Schwarzschild}, M. 1975, \apj, 195, 137

\bibitem[{{Stanishev} {et~al.}(2002){Stanishev}, {Kraicheva}, {Boffin}, \&
  {Genkov}}]{2002A&A...394..625S}
{Stanishev}, V., {Kraicheva}, Z., {Boffin}, H.~M.~J., \& {Genkov}, V. 2002,
  \aap, 394, 625

\bibitem[{{Telting} {et~al.}(2006){Telting}, {Schrijvers}, {Ilyin},
  {Uytterhoeven}, {De Ridder}, {Aerts}, \& {Henrichs}}]{2006A&A...452..945T}
{Telting}, J.~H., {Schrijvers}, C., {Ilyin}, I.~V., {et~al.} 2006, \aap, 452,
  945

\bibitem[{{Templeton} \& {Karovska}(2009)}]{2009ApJ...691.1470T}
{Templeton}, M.~R. \& {Karovska}, M. 2009, \apj, 691, 1470

\bibitem[{{Timmer} \& {Koenig}(1995)}]{1995A&A...300..707T}
{Timmer}, J. \& {Koenig}, M. 1995, \aap, 300, 707

\bibitem[{{Underhill} \& {Gilroy}(1990)}]{1990ApJ...364..626U}
{Underhill}, A.~B. \& {Gilroy}, K.~K. 1990, \apj, 364, 626

\bibitem[{{Vaughan} {et~al.}(2003){Vaughan}, {Edelson}, {Warwick}, \&
  {Uttley}}]{2003MNRAS.345.1271V}
{Vaughan}, S., {Edelson}, R., {Warwick}, R.~S., \& {Uttley}, P. 2003, \mnras,
  345, 1271

\bibitem[{{Ventura} {et~al.}(2008){Ventura}, {D'Antona}, \&
  {Mazzitelli}}]{2008Ap&SS.316...93V}
{Ventura}, P., {D'Antona}, F., \& {Mazzitelli}, I. 2008, \apss, 316, 93

\bibitem[{{Walker} {et~al.}(2005){Walker}, {Kuschnig}, {Matthews}, {Reegen},
  {Kallinger}, {Kambe}, {Saio}, {Harmanec}, {Guenther}, {Moffat}, {Rucinski},
  {Sasselov}, {Weiss}, {Bohlender}, {Bo{\v z}i{\'c}}, {Hashimoto},
  {Koubsk{\'y}}, {Mann}, {Ru{\v z}djak}, {{\v S}koda}, {{\v S}lechta}, {Sudar},
  {Wolf}, \& {Yang}}]{2005ApJ...623L.145W}
{Walker}, G.~A.~H., {Kuschnig}, R., {Matthews}, J.~M., {et~al.} 2005, \apjl,
  623, L145

\bibitem[{{Wang} {et~al.}(2008){Wang}, {Townsley}, {Feigelson}, {Broos},
  {Getman}, {Rom{\'a}n-Z{\'u}{\~n}iga}, \& {Lada}}]{2008ApJ...675..464W}
{Wang}, J., {Townsley}, L.~K., {Feigelson}, E.~D., {et~al.} 2008, \apj, 675,
  464

\bibitem[{{Zdravkov} \& {Pamyatnykh}(2008)}]{2008JPhCS.118a2079Z}
{Zdravkov}, T. \& {Pamyatnykh}, A.~A. 2008, Journal of Physics Conference
  Series, 118, 012079

\end{thebibliography}

\end{document}